\documentclass[%
 reprint,
 amsmath,amssymb,
 aps,
]{revtex4-2}

\usepackage{graphicx}
\usepackage{dcolumn}
\usepackage{bm}

\usepackage{mathtools}
\newcommand{\reals}{\mathbb{R}}
\newcommand{\integers}{\mathbb{Z}}
\newcommand{\nats}{\mathbb{N}}

\newcolumntype{P}[1]{>{\centering\arraybackslash}p{#1}}

\newcommand{\twistB}{2.291}
\newcommand{\twistBE}{0.019}
\newcommand{\twistN}{0.614}
\newcommand{\twistNE}{0.079}

\newcommand{\linknnBfour}{2.2989}
\newcommand{\linknnBEfour}{0.0009}
\newcommand{\knnBfour}{2.2988}
\newcommand{\knnBEfour}{0.0007}
\newcommand{\knnNfour}{0.634}
\newcommand{\knnNEfour}{0.014}

\newcommand{\linknnBfouralt}{2.2996}
\newcommand{\linknnBEfouralt}{0.0008}
\newcommand{\knnBfouralt}{2.2998}
\newcommand{\knnBEfouralt}{0.0007}
\newcommand{\knnNfouralt}{0.638}
\newcommand{\knnNEfouralt}{0.013}

\newcommand{\linknnBfive}{2.3696}
\newcommand{\linknnBEfive}{0.0012}
\newcommand{\knnBfive}{2.3697}
\newcommand{\knnBEfive}{0.0011}
\newcommand{\knnNfive}{0.634}
\newcommand{\knnNEfive}{0.028}

\newcommand{\linknnBsix}{2.4277}
\newcommand{\linknnBEsix}{0.0008}
\newcommand{\knnBsix}{2.4276}
\newcommand{\knnBEsix}{0.0008}
\newcommand{\knnNsix}{0.666}
\newcommand{\knnNEsix}{0.016}

\begin{document}

\preprint{}

\title{Probing center vortices and deconfinement in $\mathrm{SU}(2)$ lattice gauge theory with persistent homology}

\author{Nicholas Sale}
\email{nicholas.j.sale@gmail.com}
\homepage{https://nicksale.github.io/}

\author{Biagio Lucini}
 \altaffiliation[Also ]{Swansea Academy of Advanced Computing, Swansea University, Bay Campus, SA1 8EN, Swansea, Wales, UK}
\affiliation{%
 Department of Mathematics, Swansea University, Bay Campus, SA1 8EN, Swansea, Wales, UK
}%

\author{Jeffrey Giansiracusa}%
\affiliation{%
 Department of Mathematical Sciences, Durham University, Upper Mountjoy Campus, Durham, DH1 3LE, UK
}%


\date{\today}

\begin{abstract}
We investigate the use of persistent homology, a tool from topological data analysis, as a means to detect and quantitatively describe center vortices in $\mathrm{SU}(2)$ lattice gauge theory in a gauge-invariant manner. We provide evidence for the sensitivity of our method to vortices by detecting a vortex explicitly inserted using twisted boundary conditions in the deconfined phase. This inspires the definition of a new phase indicator for the deconfinement phase transition. We also construct a phase indicator without reference to twisted boundary conditions using a simple $k$-nearest neighbours classifier. Finite-size scaling analyses of both persistence-based indicators yield accurate estimates of the critical $\beta$ and critical exponent of correlation length $\nu$ of the deconfinement phase transition.
\end{abstract}

\maketitle

\parskip=10pt


\section{Introduction}
\label{sec:Introduction}

Quantum chromodynamics (QCD) poses several outstanding problems in particle physics, including the mechanism of confinement and the deconfinement phase transition, mass-gap generation, and chiral symmetry breaking \cite{greensite2011introduction}. These phenomena are non-perturbative and are therefore typically investigated through the framework of Lattice QCD. Supplementing this traditional approach, there is an emerging body of work exploring the use of machine learning and data analysis tools in extending Monte Carlo analysis of Lattice QCD towards generating the insights required to tackle these open problems. References include \cite{PhysRevB.96.184410, PhysRevD.103.014509, https://doi.org/10.48550/arxiv.2111.05216} among others. For example, one application of machine learning is to classify phases of the theory based on sampled configurations, learning observables that function as order parameters for the phase transitions undergone by QCD, following the quantitative programme outlined for spin models in~\cite{Giannetti:2018vif}. 

Focusing on confinement, a compelling potential mechanism relies on the presence of topological defects called center vortices in confining gauge configurations \cite{THOOFT19781, CORNWALL1979392, tHooft:1979rtg}. Vortex-like configurations have been shown to exist in pure gauge theories (e.g.,~\cite{DelDebbio:1996lih,Faber:1997rp,DelDebbio:1998luz,Bertle:1999tw,Faber:1999gu,Engelhardt:1999fd,deForcrand:1999our,Engelhardt:1999xw,Engelhardt:2000wc,Bertle:2000qv,Langfeld:2001cz,Greensite:2003bk,Bruckmann:2003yd,Engelhardt:2003wm,Boyko:2006ic,Ilgenfritz:2007ua,Bornyakov:2007fz,OCais:2008kqh,Engelhardt:2010ft,Bowman:2010zr,OMalley:2011aa,Trewartha:2015ida,Trewartha:2015nna,Greensite:2016pfc,Trewartha:2017ive,Biddle:2018dtc,Spengler:2018dxt}) and have recently been observed in lattice simulations of QCD~\cite{Biddle:2022zgw}. While order parameters have been constructed for the confinement-deconfinement phase transitions in Yang-Mills theories that are based on the topological symmetry related to the conservation of the number of vortices (see, e.g.,~\cite{deForcrand:1999our,DelDebbio:2000cx,DelDebbio:2000cb}), identification of vortices proves to be a more challenging undertaking. In fact, existing methodologies for exposing vortices in gauge theories rely on performing gauge fixing and projection \cite{universe7050122}. This procedure suffers from the problem of Gribov ambiguities \cite{GRIBOV19781,STACK2001529}, which, even with careful choices of the gauge fixing condition (for instance, following the prescription of~\cite{deForcrand:2000pg}), can be mitigated only in part. Motivated by the physical appeal of a fully gauge independent description, we investigate the possibility of instead analysing vortices and performing phase classification in a gauge-invariant manner by making use of persistent homology~\cite{Edelsbrunner2002TopologicalPA} alongside other tools from topological data analysis, a novel approach to data (in our case, consisting of gauge configurations) that places emphasis on rigorous classifications of shapes of datasets based on their topological properties. Rather than considering full QCD, in this paper we instead develop a methodology for the pure gauge $\mathrm{SU}(2)$ lattice gauge theory, which also exhibits a deconfinement transition potentially driven by center vortices. In the context of vortex identification, this system represents a toy model of QCD where the quarks have been removed and the gauge group is simplified from $\mathrm{SU}(3)$.

Among other previous work on the use of persistent homology and computational topology to investigate systems from physics \cite{PhysRevE.98.012318, Hirakida2018PersistentHA, tran2020topological, olsthoorn2020finding, cole2020quantitative, Donato2016PersistentHA, spitz2020finding, sale2022quant, ARIASTAMARGO2022137376,PhysRevD.105.066002,PhysRevB.106.054210} (a useful survey is \cite{https://doi.org/10.48550/arxiv.2206.15075}), we remark on two that look at QCD and lattice gauge theories in particular. Kashiwa, Hirakida and Kouno noted the sign problem in simulating dense QCD and instead considered an effective model in the form of a modified 3D Potts model where each $\integers_3$ Potts spin corresponds to a Polyakov loop \cite{https://doi.org/10.48550/arxiv.2103.12554}, avoiding the need to construct a filtered complex from gauge fields directly. Computing the persistence of pointclouds of lattice sites sharing the same Potts spin, they were able to probe the phase structure of the model using the average and maximum birth-death ratio of points in the resulting persistence diagrams. More recently, Sehayek and Melko investigated the 2D and 3D $\integers_2$ lattice gauge theories \cite{Sehayek}. Given a configuration of the model, they place a point at the centre of each spin-down lattice link and compute the Vietoris-Rips filtered complex of the resulting pointcloud. The $\beta_1$ Betti curve then provides the number of closed strings of down spins and their sizes. These quantities and the filtered complex are gauge variant, but by averaging the loop count over many configurations they indirectly measure the density of vison defects which produces a clear indicator of the phase transition in the 3D model when plotted as a function of temperature.

Our main contributions are as follows:
\begin{itemize}
    \item We introduce a gauge-invariant construction that produces a filtered complex (for input to persistent homology) from a field configuration for $\mathrm{SU}(2)$ lattice gauge theory, and we argue that this pipeline should see center vortices.
    \item We demonstrate that the resulting persistent homology is able to detect an explicitly inserted vortex in the deconfined phase by showing that it distinguishes configurations generated using twisted boundary conditions. In particular, we show that a phase indicator can be recovered by comparing the $PH_2$ persistence diagrams of configurations generated using twisted boundary conditions and configurations sampled using the usual Wilson action, similarly to the vortex free energy order parameter. This phase indicator allows us to estimate the critical $\beta$ and critical exponent of correlation length via finite-size scaling.
    \item Using a $k$-nearest neighbours classification of persistence images, we identify the deconfinement phase transition directly from configurations generated using the usual Wilson action. Moreover we accurately estimate the critical $\beta$ and critical exponent of correlation length via finite-size scaling.
\end{itemize}

The rest of this paper is organised as follows. In Section \ref{sec:SU(2)} we introduce the $\mathrm{SU}(2)$ lattice gauge theory, the deconfinement phase transition and the center vortex picture for confinement. In Section \ref{sec:Methods} we review the techniques we use including persistent homology, $k$-nearest neighbours classification and finite-size scaling analysis. In Section \ref{sec:Analysis} we introduce two different persistent homology-based phase indicators for deconfinement, meaning statistics that are zero in one phase and non-zero in the other, which we use to quantitatively analyse the phase transition. Finally in Section \ref{sec:Conclusions} we discuss our findings and identify potential directions for future work. The appendices contain more detailed reviews of some of the tools we use.

The data and the code used for this work are being released respectively in Refs~\cite{datarepo}~and~\cite{softwarerepo}.

\section{$\mathrm{SU}(2)$ Lattice Gauge Theory}
\label{sec:SU(2)}

\subsection{The Model}

A configuration of the 4D $\mathrm{SU}(2)$ lattice gauge theory is specified by $\mathrm{SU}(2)$-valued variables $U_\mu(x)$ located on each link $(x, \mu)$ of an $N_t \times N_s^3$ lattice $\Lambda$ with periodic boundary conditions, where $\mu \in \{ 0, 1, 2, 3 \}$ describes the direction in which the link emanates from the lattice site $x \in \Lambda$. In practice, $U_\mu(x)$ lies in the fundamental representation of $\mathrm{SU}(2)$, taking the form of a $2 \times 2$ complex matrix. To simulate at non-zero temperature we ensure $N_t \ll N_s$. The gauge symmetry is generated by gauge transformations $\Omega(x) \in \mathrm{SU}(2)^\Lambda$ sending each $U_\mu(x) \mapsto \Omega^\dag(x) U_\mu(x) \Omega(x + \hat\mu)$, where $x + \hat\mu$ denotes the lattice site one step in the $\mu$ direction from $x$. The observables that are invariant under these transformations are traces of products of the link variables along closed paths $C$, also known as Wilson loops $W(C)$. The simplest non-trivial example is the Wilson loop around a $1 \times 1$ plaquette $(x, \mu, \nu)$ of the lattice:
$$W_{\mu, \nu}(x) = \frac{1}{2} \, tr \Big[ U_\mu(x) \, U_\nu(x + \hat{\mu}) \, U_\mu^\dagger(x + \hat{\nu}) \, U_\nu^\dagger(x) \Big].$$
We use this to define the Wilson action given a configuration $\mathbf{U} = \{ U_\mu(x) \}_{(x,\mu)}$ as
\begin{equation}
   S(\mathbf{U}) = - \frac{\beta}{4} \sum_{x, \mu < \nu} W_{\mu, \nu}(x) \label{eqn:action}
\end{equation}
where $\beta = 4/g^2$ and $g$ is the gauge coupling parameter. This in turn allows us to define the vacuum expectation value of any given observable $A(\mathbf{U})$ as
\begin{equation}
\langle A \rangle = \frac{\int d\mathbf{U} \, A(\mathbf{U}) \, e^{-S(\mathbf{U})}}{\int d\mathbf{U} \, e^{-S(\mathbf{U})}} \label{eqn:vev}
\end{equation}
where $d\mathbf{U} = \prod_{x,\mu} dU_\mu(x)$ is a product of Haar measures over $\mathrm{SU}(2)$ for each link variable. In practice we estimate expectations using Monte Carlo methods, where Eq. \eqref{eqn:vev} becomes a simple mean of the observed values.

\subsection{Deconfinement and Center Vortices}

The model introduced above exhibits two phases \--- a confined phase at low $\beta$ and a deconfined phase at high $\beta$ \--- and the phase transition between these is known as the deconfinement transition. This transition occurs only in the spatial continuum limit $N_s \rightarrow \infty$ and while we care also about the temporal limit $N_t \rightarrow \infty$, it's worth noting that a critical $\beta$ is well defined for each finite $N_t$ and it is these we will be estimating. Confinement in the $\mathrm{SU}(2)$ lattice gauge theory can be characterised in a number of ways \cite{greensite2011introduction}, including:
\begin{itemize}
    \item \textbf{Area law for Wilson loops.} Let $W(C)$ denote the value of a Wilson loop around a closed curve $C = R \times T$, consisting of rectangle of edges $R$ in a space-like direction and $T$ along the temporal direction. We consider the limit of large area $\mathcal{A}(C) = R T$. In the confined phase at low $\beta$ we have that $\langle W(C) \rangle \propto exp(-\sigma \mathcal{A}(C))$ where $\sigma$ is known as the string tension. In the deconfined phase at high $\beta$ we have that $\langle W(C) \rangle$ decays exponentially instead with $\mathcal{P}(C)$, the perimeter of $C$.
    \item \textbf{Vanishing Polyakov loop.} Define the Polyakov loop at a point $x$ in the lattice as
    $$P(x) = \frac{1}{2} \, tr\Big[ U_0(x) \, U_0(x + \hat{0}) \, \dots \, U_0(x - \hat{0}) \Big]$$
    (note this only depends on the spatial coordinates of $x$). This represents a Wilson loop that has a non-zero winding number around the time direction of the periodic lattice. In the confined phase we have $\langle P(x) \rangle = 0$ whereas in the deconfined phase we have $\langle P(x) \rangle \neq 0$.
\end{itemize}

Several possible pictures of what drives the deconfinement transition, in both this model and QCD, have been proposed. Here we focus on the center vortex picture~\cite{THOOFT19781, CORNWALL1979392}. Fix a time slice at time $t$. Given two closed oriented contours $C$ and $C^{\prime}$ in that 3-dimensional slice with linking number $m$, a loop operator $B(C^{\prime}, t)$ can be defined that has the following commutation algebra with the Wilson loop $W(C, t)$: 
\begin{equation}
    W({C}, t) B({C^{\prime}}, t) - (-1)^m B({C^{\prime}}, t) W({C}, t) = 0.
\end{equation}
This equation defines the 't Hoof algebra~\cite{THOOFT19781}. For simplicity, we consider planar non-intersecting curves ${C}$ and ${C^{\prime}}$, for which $m = 0,1$. The operator $B(C^{\prime}, t)$ is called the 't Hooft loop. When acting on a gauge configuration, $B({C^{\prime}}, t)$ creates a magnetic flux with the resulting observable effect of multiplication by -1 of all Wilson loops having support on curves ${C}$ with linking number 1 with ${C^{\prime}}$. For this reason, the 't Hooft loop is said to be a vortex creation operator. Since the center of the group, which in our case is
$Z(\mathrm{SU}(2)) = \{ I, -I \} \cong \integers_2$, plays a role in the 't Hooft algebra (as exposed by the factor $(-1)^m$), the vortices created by the 't Hooft loop operator are called center vortices. Fixing the curve $C^\prime$ for all time slices $t$, we see that a vortex traces out a surface in 4-space, closed by the periodic boundary conditions.

In the limit of weak fields, where the theory is deconfined, all Wilson loops are close to unity. Confining configurations are expected to have Wilson loops that largely deviate from unity. In particular, Wilson loops close to -1 can be obtained from a weak field configuration through the injection of center vortices generated with appropriate insertions of 't Hooft loop operators. Moving from this observation, operationally we can define a center vortex to be a collection of plaquettes in the dual lattice (in the sense of the dual graph) that form a closed surface (with the closedness being a consequence of the Bianchi identities) and that carry a non-trivial charge in  $\integers_2$, corresponding to the $-1$ element. To carry a non-trivial center charge means that any Wilson loop in the lattice that topologically links with this surface is multiplied by that charge. 

In the confined phase center vortices are found to form large surfaces, often wrapping round the periodic boundaries, that percolate throughout the lattice \cite{PhysRevD.61.054504}. Therefore, given a particular Wilson loop $W(C)$, the number of vortices that link with $C$ is proportional to the enclosed area $\mathcal{A}(C)$, leading to the area law for the suppression of $\langle W(C) \rangle$. In the deconfined phase, the center vortices become smaller and more sparse, ensuring that for sufficiently large Wilson loops, only those vortices close to the curve $C$ have a chance of linking with it, leading to the perimeter law. Similarly we see that only in the confined phase, where vortices may wrap around the periodic boundary conditions of the lattice, is there a chance they may link with a Polyakov loop, suppressing its expectation. For an overview of the evidence supporting the center vortex picture see \cite{greensite_evidence}.
In practice, center vortices generated by the system have some finite thickness, so that only larger Wilson loops may fully link with them and obtain the full center charge. Loops that partially link may still obtain a partial charge, some factor lying between $I$ and $-I$ in $\mathrm{SU}(2)$. 

While the concepts of vorticity and creation of a vortex through the insertion of a 't Hooft loop are well understood in terms of symmetry and boundary conditions in a finite volume (see for instance~\cite{THOOFT19781,deForcrand:1999our,DelDebbio:2000cx,DelDebbio:2000cb}), quantum fluctuations make vortex identification a much more involved process, with currently used prescriptions not fully validated from first principles. A widely used method to detect and analyse these thick vortices is to transform configurations to the maximal center gauge, where each matrix $U_\mu(x)$ is as close to either $I$ or $-I$ as possible, then project the matrices onto whichever of $I$ or $-I$ is closer. After projection, the Wilson loops of plaquettes take values either $1$ or $-1$ and the latter are identified as projected vortices, or P-vortices. It has been shown that the locations of these correlate with the unprojected thick vortices \cite{Faber_2001}. However, the projection means that we lose gauge invariance, as well as geometric information such as the thickness of the vortices. In this work we introduce a method designed to look for thin vortices, but we argue that, through the use of persistent homology, the results may tell us something about these thick vortices too.

There is no analytical formula for the critical value of $\beta$ for each value of $N_t$, but detailed various numerical studies have shown good agreement with each other. We will compare our results to those found in \cite{lucini_SU(N)_transition} which we reproduce in Table \ref{tab:crit_betas}. We also estimate the critical exponent of correlation length $\nu$ which, as a consequence of the Svetitsky–Yaffe conjecture \cite{SVETITSKY1982423}, is the same as for the 3D Ising model. We therefore compare our measurement of $\nu$ to the high precision estimate $\nu = 0.629971(4)$ from \cite{precision_ising}.

\begin{table}[h!]
\centering
\begin{tabular}{ || w{c}{4em} | w{c}{6em} || } 
 \hline
 $N_t$ & $\beta_c$ \\
 \hline
 4 & 2.2986(6) \\ 
 5 & 2.37136(54) \\
 6 & 2.4271(17) \\
 \hline
\end{tabular}
\caption{Estimates for the critical value of $\beta$ for the deconfinement phase transition in the $\mathrm{SU}(2)$ lattice gauge theory for the values of $N_t$ we consider in this paper. Reproduced from \cite{lucini_SU(N)_transition}.}
\label{tab:crit_betas}
\end{table}

\subsection{Twisted Boundary Conditions}
\label{sec:TwistedBoundary}

\begin{figure}[ht]
    \centering
    \scalebox{0.25}{\includegraphics{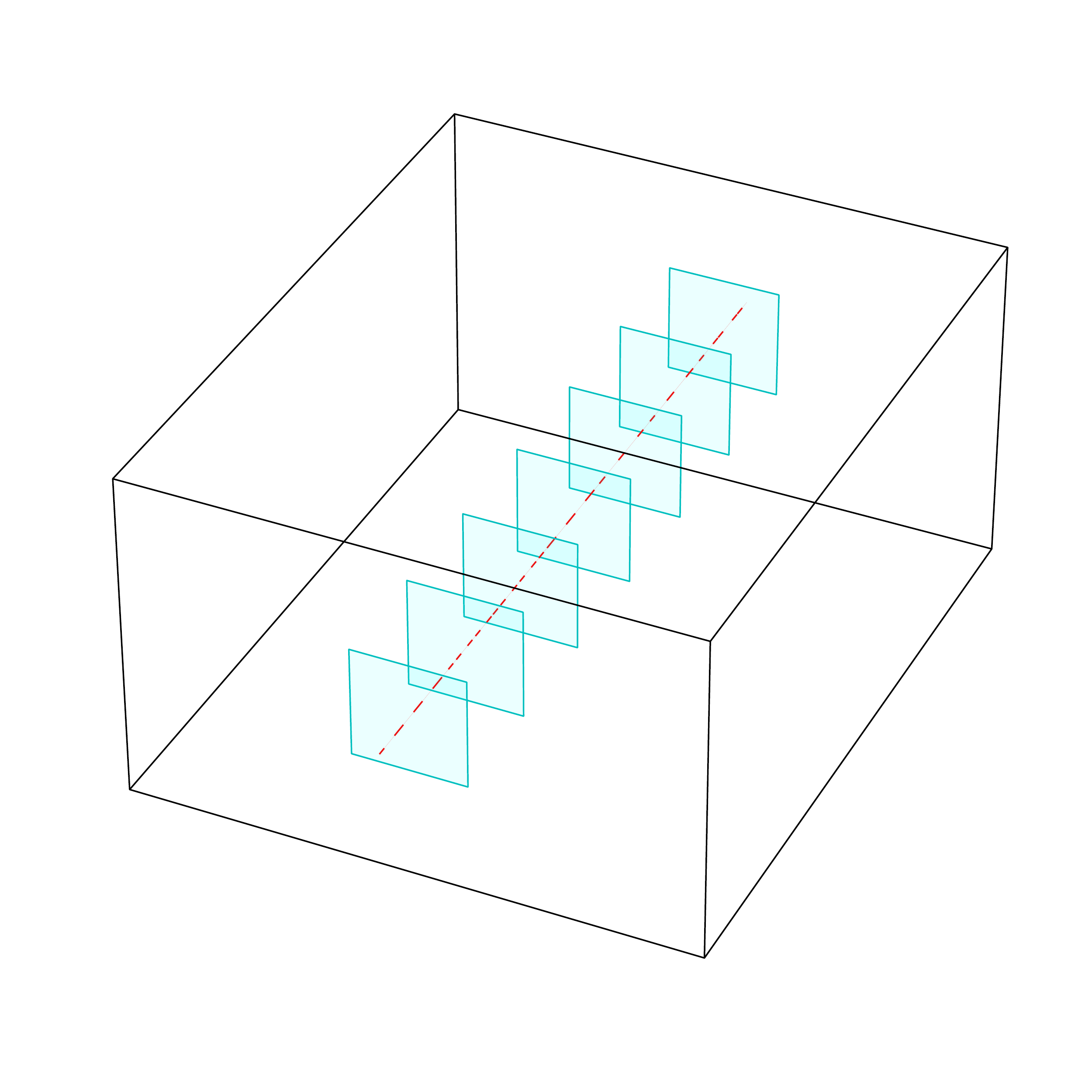}}
    \caption{A lower dimensional illustration of a co-closed collection of plaquettes that wraps around the periodic boundary conditions of the lattice. Plaquettes in a 3D lattice link with edges in the dual lattice, so the condition of being co-closed means that the collection of those linking edges forms a closed loop. In this case the loop is closed by the periodic boundary conditions. Going to 4 dimensions, we imagine repeating the co-closed line of plaquettes along the new dimension, forming a co-closed surface of plaquettes.}
    \label{fig:tbc}
\end{figure}

In order to test the sensitivity of our method to thin vortices we will make use of the trick of imposing twisted boundary conditions \cite{THOOFT1979141}. The idea is that we choose some co-closed collection of plaquettes in the lattice, i.e., plaquettes that link with some closed surface in the dual lattice, and negate their contribution to the action. See Figure \ref{fig:tbc} for an illustration of co-closed collections of plaquettes. In our case we choose the plaquettes
$$T = \{ (x, \mu, \nu) = ((0,0,y,z), 0, 1) \,\, \vert \,\, 0 \leq y, z < N_s \}$$
corresponding to a surface wrapping round the latter two spatial dimensions of the lattice. The action with twisted boundary conditions becomes
\begin{equation}
    S_T(\mathbf{U}) = - \frac{\beta}{4} \Bigg[\sum_{\substack{x,\mu < \nu \\ (x, \mu, \nu) \not\in T}} W_{x, \mu, \nu} - \sum_{\substack{x,\mu < \nu \\ (x, \mu, \nu) \in T}} W_{x, \mu, \nu}\Bigg]
    \label{eqn:tbc_action}
\end{equation}
which we refer to as the twisted action.

This modification of the action allows the lattice to support an odd number of center vortices wrapping in the $yz$ plane, which is prohibited by the usual periodic boundary conditions of the Wilson action. It is important to note that we are talking about the boundary conditions of the gauge field on the lattice and not the lattice itself. We are not twisting the lattice and forming any kind of M{\"o}bius band, rather it is the gauge field which obtains a factor of $-I$ as we loop around the lattice. We can alternatively think of this twisted action as explicitly inserting a thin vortex into the system on the surface defined by $T$, so that the system is forced to generate a (thick) vortex to cancel it out. We shall denote expectations calculated with respect to this twisted action by $\langle A \rangle_{\text{twist}}$, where $A$ is a generic observable. 

Twisted boundary conditions give us an alternative way to characterise confinement and the deconfinement transition. Magnetic and electric flux free energies can be defined in terms of the ratio of partition functions for the twisted and Wilson actions and the behaviour of these can be shown to imply the area law decay for the Wilson loop \cite{Tomboulis1985} and therefore confinement. 

\section{Methods}
\label{sec:Methods}

\subsection{Background on Persistent Homology}
\label{sec:BackgroundPH}
Persistent homology is a computational topology tool introduced in its modern form in \cite{Edelsbrunner2002TopologicalPA} and popularised in \cite{Carlsson2009TopologyAD}. It is one of the main tools of Topological Data Analysis. We shall give a brief overview here following that in \cite{sale2022quant}, but for a more complete review of persistent homology useful references are \cite{carlsson2020persistent, ph_survey_edels_harer, otter, ghrist_barcodes}.

Given a topological space, such as a manifold or a simplicial/cubical complex, homology is an algebraic way of describing the 'holes' in the space. In particular, the spaces we consider will be cubical complexes. A very brief technical introduction to cubical complexes and their homology can be found in Appendix \ref{appendix:cubical}. In general terms, the $k$\textsuperscript{th} homology $H_k(C)$ of a cubical complex $C$, computed with coefficients in a field (here $\integers_2$), is a vector space that has a basis in $1$-$1$ correspondence with the $k$-dimensional holes in $C$. As an example, consider removing a unit cube from $\reals^3$. This leaves behind a 2-dimensional hole since it is enclosed by a 2-dimensional surface and as such would be recorded by $H_2$. Given a map of cubical complexes $f : C \rightarrow C^\prime$ (e.g. an inclusion map), we obtain induced linear maps $f_k: H_k(C) \rightarrow H_k(C^\prime)$. The rank of $f_k$ tells us how many of the $k$-dimensional holes survived after being mapped into $C^\prime$ i.e. how many \textit{persisted}. Given some data $D$, the idea of (cubical) persistence then is to construct a collection of cubical complexes indexed by the reals
$$F_D : \reals \rightarrow \text{CubicalComplex}$$
using the data, so that we have inclusions
$$F_D(s) \subseteq F_D(t)$$
for all $s \leq t \in \reals$. We call such an $\reals$-indexed collection of complexes and inclusion maps a filtered complex. Since the $F_D(s)$ are each subcomplexes of the final complex $F_D(\infty)$, we can specify the filtered complex by assigning to each cube in the final complex the index at which it first appears, and then $F_D(s)$ is the subcomplex consisting of all cubes that have appeared at or before $s$.

Applying homology now yields an $\reals$-indexed collection of vector spaces
$$H_k \circ F_D : \reals \rightarrow \text{VectorSpace}$$
and linear maps
$$H_k(F_D(s)) \rightarrow H_k(F_D(t))$$
and we may use the ranks of these maps to identify when new holes are born, how long they persist through the filtered complex, and when they die. We summarise this information as a multi-set called a persistence diagram $PH_k(F_D) \subset \{ (a,b) \in \reals \times (\reals \cup \{ \infty \}) \mid a \leq b \}$ that contains a pair $(b,d)$ every time a $k$-dimensional hole is born in $F_D(b)$ and dies in $F_D(d)$. In the case that a hole persists even in the final complex $F_D(\infty)$, we write $d = \infty$. We shall make use of such points later. It is said that a feature is born at $b$, dies at $d$ and that its persistence is $d-b$. This can also be represented as a barcode (a multi-set of intervals $[b, d)$). There are a few ways to define distances between persistence diagrams, but those which are most commonly used are the bottleneck and Wasserstein distances (for details see e.g., \cite{books/daglib/0025666}). For many typical choices of filtered complex a small change in the input data $D$ leads to only a small change in the persistence diagram $PH_K(F_D)$ as measured by these distances. This property of persistent homology is known as stability, and makes persistence a useful tool for dealing with real-world, noisy data. 

In this work $D$ is a single configuration of the $\mathrm{SU}(2)$ lattice gauge theory. We therefore obtain a persistence diagram for each sampled configuration and we can consider statistics computed from these diagrams.

In their form as multi-sets, persistence diagrams do not lend themselves to use directly as inputs for many standard machine learning models and can contain many points, taking up a large amount of computer memory. However there are numerous methods to represent persistence diagrams as fixed-size vectors. In this work we make use of a particular vectorisation called a persistence image \cite{persistenceImages}. Let $g_{\alpha,\beta} : \reals^2 \rightarrow \reals$ denote a 2D Gaussian of standard deviation $\sigma$ centered at $(\alpha,\beta)$: $$g_{\alpha,\beta}(x,y) = \frac{1}{2\pi\sigma^2}exp\bigg[-\frac{(x-\alpha)^2 + (y-\beta)^2}{2\sigma^2}\bigg].$$ Given a persistence diagram $PH_k = \{ (b_i, d_i) \}_{i \in I}$, its persistence surface is the function $\rho_k : \reals^2 \rightarrow \reals$ obtained by translating each point $(b,d) \in PH_k$ with $d \neq \infty$ into birth-persistence coordinates $(b, d-b)$, then placing Gaussians with variance $\sigma^2$ on them, weighted by the persistence of the point:
$$\rho_k(x,y) = \sum_{(b,d)\in PH_k}(d-b) \, g_{b,d-b}(x,y).$$
The $(n_I)^2$-dimensional persistence image $PI_k$ is obtained by discretizing a rectangular region of the domain of $\rho_k$ into a collection of $n_I \times n_I$ pixels $p_i$ and integrating $\rho_k$ within each:
$$PI_k^i = \iint_{p_i} \rho_k(x,y)dxdy.$$

So long as we choose the same $\sigma$ and discretization for each diagram, we can compute averages and variances component-wise. Besides emphasising high-persistence points, the linear weighting by the persistence ensures the stability of the persistence image. Finally we note that, as discussed in \cite{persistenceImages}, machine learning models trained on persistence images are generally insensitive to the resolution and variance parameters $n_I$ and $\sigma$. We also found this to be the case in our previous work \cite{sale2022quant}. Therefore in this work, we shall fix the parameters with a resolution of $25 \times 25$ and $\sigma$ equal to $5\%$ of a pixel.

\subsection{Filtered Complex}
\label{sec:filtration}

As described in the previous section, to apply persistent homology we must choose how to define a filtered complex for a given configuration $\boldsymbol{U} = \{U_\sigma(x)\}$. We present a filtered complex $F_{\boldsymbol{U}}$ which is constructed based on Wilson loops and which will therefore give gauge-invariant persistence diagrams.

The idea is to explicitly construct a cubical model of vortex surfaces, under the assumption that vortices are thin. Note that since we have periodic boundary conditions, spacetime in this model forms a 4-torus $S^1 \times S^1 \times S^1 \times S^1$. The lattice $\Lambda$ defines a decomposition of the spacetime manifold into a cubical complex $X$ in which there is a vertex for each lattice site, an edge for each link in the lattice, a 2-cube for each plaquette, etc.  The dual lattice $\Lambda^*$  determines a dual cubical complex decomposition $Y$ of spacetime. The vertices of $Y$ are the centres of the 4-cubes of $X$. More generally, the $d$-cubes of $Y$ are in bijection with the $(4-d)$-cubes of $X$, where the bijection pairs a cube in $X$ with the unique cube in $Y$ for which the intersection is a single point.

Wilson loops live on the cubical complex $X$, while vortex sheets live on the dual complex $Y$.  We will construct a filtration of $Y$ by letting each 2-cube enter at a filtration index given by the value of the Wilson loop around the boundary of the dual 2-cube in $X$.

In more detail, denote by $c_A(y) = \prod_{\mu \in A}[y, y + \overrightarrow\mu]$ a cube in $Y$, where $[y, y + \overrightarrow\mu]$ is the line segment between lattice site $y \in \Lambda^*$ and $y + \overrightarrow\mu$. The cube $c_\emptyset(y)$ is just the point $y$ itself. The dimension $d$ of $c_A(y)$ is $\vert A \vert$ and we will refer to it as a $d$-cube. The boundary $\partial c_A(y)$ of a $d$-cube is the set of its $d-1$-cube faces. For example $\partial c_{\{\mu\}}(y) = \{c_\emptyset(y), c_\emptyset(y+\overrightarrow\mu)\}$. With this notation, our observation of the bijection between plaquettes in $X$ and $Y$ becomes that the 2-cube $c_{\{\mu,\nu\}}(y)$ is matched with the 2-cube $c_{\{\sigma, \tau\}}(x)$ in $X$ (defined similarly) that is used to define the Wilson loop $W_{\sigma, \tau}(y + \overrightarrow\mu + \overrightarrow\nu)$ where $\{\sigma, \tau \} \cap \{ \mu , \nu \} = \emptyset$.

To define the filtered complex we will give a filtration index $f(c_A(y)) \in \reals$ for each cube $c_A(y)$ in $Y$ specifying when it appears. Then $F_{\boldsymbol{U}}(s)$ is the subcomplex of $Y$ consisting of all cubes $c$ for which $f(c) \leq s$. That is, $$F_{\boldsymbol{U}}(s) = f^{-1}(-\infty, s].$$ Since we are attempting to model vortex surfaces, we will initially specify when the 2-cubes are to enter the filtered complex and then introduce the cubes of other dimensions based on these.

\begin{figure}[ht]
    \centering
    \scalebox{0.17}{\includegraphics{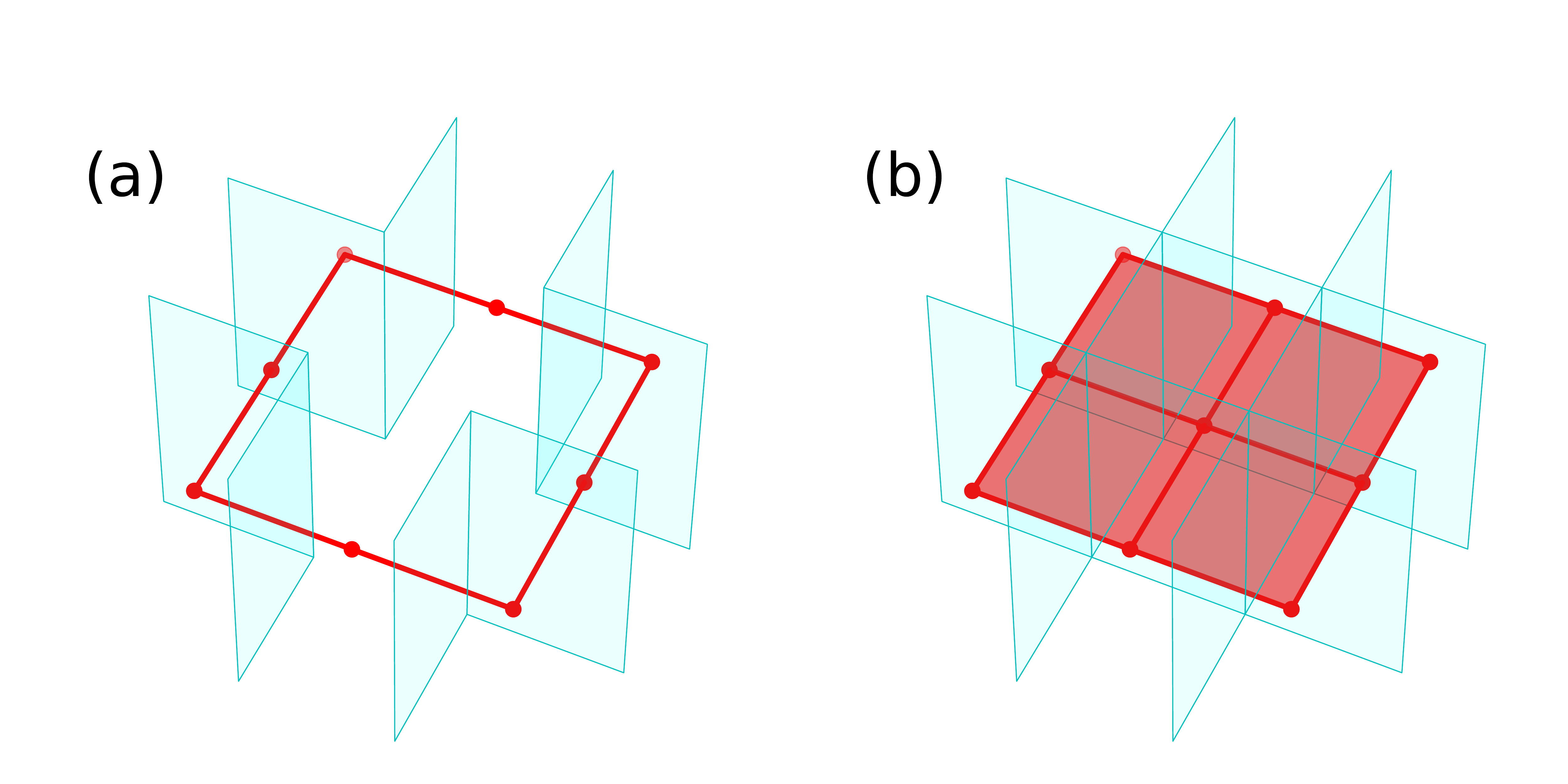}}
    \caption{A lower dimension illustration of the idea behind the filtered complex. In 3 dimensions, center vortices form closed 1-dimensional curves that link with 2-dimensional plaquettes. In this setting we would include edges (dark/red) according to the Wilson loop around the plaquette (light/cyan) they intersect. (a) Early on in the filtered complex we include edges that link with plaquettes with negative Wilson loop values. In this way we build explicit cubical models of 1-dimensional center vortices which are then detected in $PH_1$. (b) Later on we eventually fill in the rest of the edges and the 2-cubes between them, destroying the $PH_1$ features corresponding to the vortices. Moving from 3 dimensions to 4, we are inserting plaquettes instead of edges and instead of a closed curve we obtain a closed surface which we detect with $PH_2$.}
    \label{fig:3dvortex}
\end{figure}

\begin{figure*}[ht]
    \centering
    \scalebox{0.4}{\includegraphics{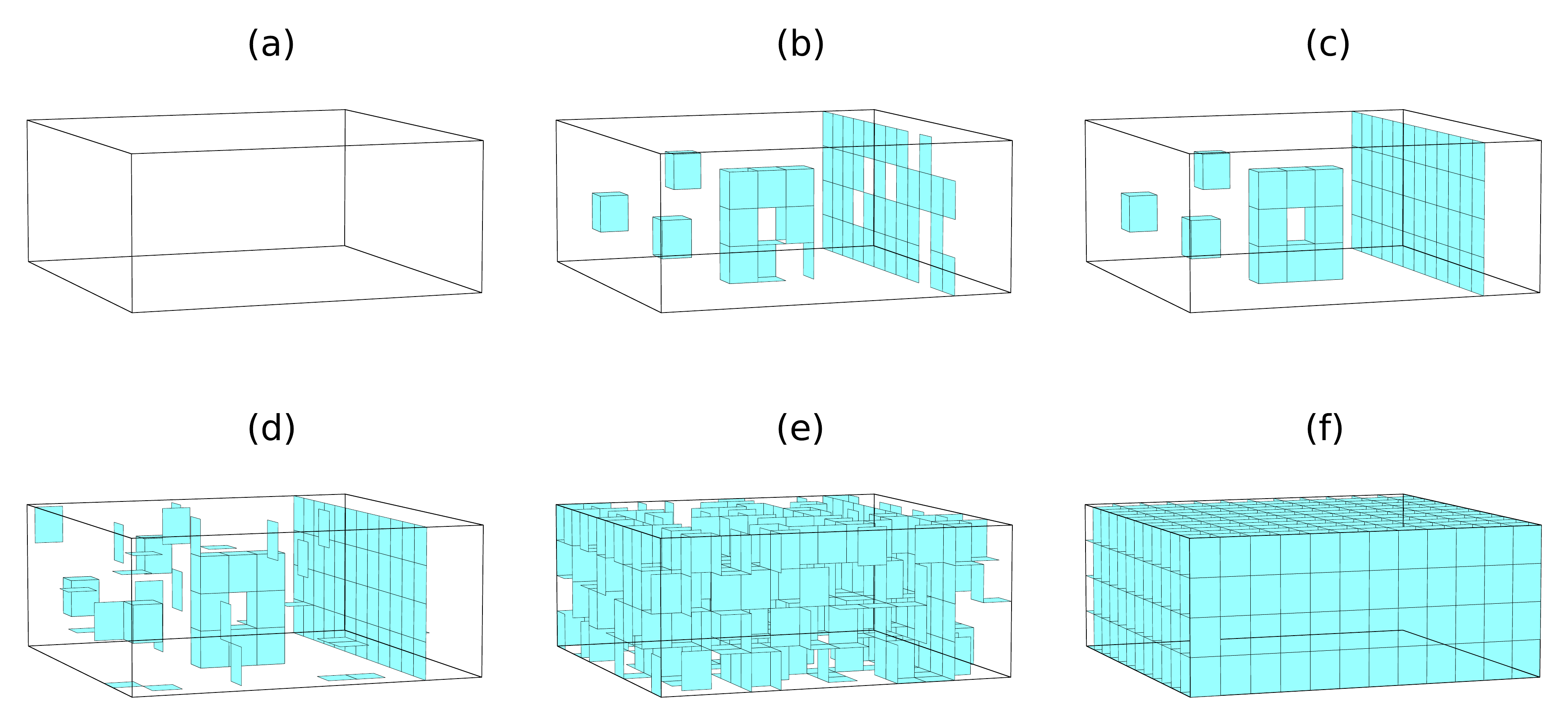}}
    \caption[A cartoon of a 3D slice of the filtered complex.]{A cartoon of a 3D slice of the filtered complex. (a) At the beginning of the filtration the complex is empty. (b) At low Wilson loop values we begin to form vortex surfaces with transient features detected in $H_0$ and $H_1$. (c) The vortex surfaces close, becoming detectable in $H_2$. (d) At higher Wilson loop values, vortex surfaces persist while other plaquettes begin to be included in the complex. (e) The vortex surfaces may become filled in if all plaquettes inside the surface are included, killing the corresponding $H_2$ class. (f) Eventually the whole 4-torus cubical complex is filled in.}
    \label{fig:vortex_filt}
\end{figure*}

Our construction of the function $f$ is the following:
\begin{enumerate}
    \item We introduce each 2-cube $c_{\{\mu , \nu\}}(y)$ in our filtered complex at index $$f(c_{\{\mu, \nu\}}(y)) = W_{\sigma, \tau}(y + \overrightarrow\mu + \overrightarrow\nu)$$ where $\{\sigma, \tau \} \cap \{ \mu , \nu \} = \emptyset$. That is, at an index equal to the value of the Wilson loop around the plaquette in $X$ paired with it by the bijection.
    \item Since a 2-cube is not allowed to be included before its constituent 1-cubes and 0-cubes in a cubical complex, we introduce these at the smallest index of all the 2-cubes they are incident to. So
    $$f(c_A(y)) = \min \{\, f(C) \,\,\vert\,\, c_A(y) \in \partial C \,\}$$
    when $\vert A \vert \leq 1$.
    \item For the 3-cubes and 4-cubes we follow a clique-like rule where we introduce a cube as soon as all of its boundary cubes are introduced. So
    $$f(c_A(y)) = \max \{\, f(C) \,\,\vert\,\, C \in \partial c_A(y) \,\}$$
    when $\vert A \vert \geq 3$.
\end{enumerate}
Thus for $s < -1$, $F_{\boldsymbol{U}}(s)$ is the empty complex and for $s \geq 1$, $F_{\boldsymbol{U}}(s)$ is the filled in tiling homeomorphic to a 4-torus. Going between these values, the first cubes to enter $F_{\boldsymbol{U}}$ are surfaces made up of plaquettes in bijection with Wilson loops that are close to $-1$. The idea therefore is that thin vortex surfaces will enter the filtered complex early. Moreover, since small Wilson loops like those considered here still pick up a partial charge from thick vortices, surfaces representing those thick vortices ought to enter the filtered complex earlier than they otherwise would have. We expect to detect these closed surfaces in persistent $H_2$ (since we compute homology with $\integers_2$ coefficients, the orientability of the surfaces does not impact this). We may also see other topological features such as the presence of handles or holes in $H_1$, as well as the transient low-persistence points in persistent $H_0$ and $H_1$ that arise as the surface forms near the start of the filtration. An illustration of the connection between Wilson loops and the inclusion of vortices into the filtration is shown in Figure \ref{fig:3dvortex}. An illustration of how to imagine what the filtered complex is aiming to do is shown in Figure \ref{fig:vortex_filt}.

It is worth noting the difference in approach from our previous work using persistent homology to identify vortices in 2D XY models \cite{sale2022quant}. Vortices there were point defects (located at vertices of the dual lattice), which we aimed to detect by constructing 1-dimensional loops in the original lattice that encircled them. Here, we are modelling the vortex surfaces in the dual lattice directly.

It is straightforward to see that this filtered complex $F_{\boldsymbol{U}}$ is stable with respect to perturbations of the $\mathrm{SU}(2)$ link variables since the Wilson loop $W_{\mu,\nu}(y)$ is a linear map due to the linearity of the trace and is therefore Lipschitz continuous. The stability property of persistent homology \cite{stability} therefore ensures that a small perturbation of the link variables only results in a small perturbation of the resulting persistence diagram with respect to the bottleneck distance.

\subsection{$k$-Nearest Neighbours Classification}
\label{sec:kNN}

In Section \ref{sec:Deconfinement} we will make use of $k$-nearest neighbours ($k$NN) classification to map the persistence images obtained from configurations onto phases as in \cite{sale2022quant}. For vector-valued data, a $k$NN classifier is a non-parametric model that models a categorical dependent variable $y(\mathbf{x}) \in \nats$, where $\mathbf{x} \in \reals^N$. The behaviour of the model is determined by the training data $\{(\mathbf{x}_i, y_i)\}$ and a choice of the hyper-parameter $k \in \nats$. Given a new input $\mathbf{x}$, it finds the $k$ indices $i^1_\mathbf{x}, \ldots, i^k_\mathbf{x}$ that minimise the Euclidean distance $\vert\vert \mathbf{x} - \mathbf{x}_i \vert\vert_2$. It then outputs the most common label among the $y_{i^1_\mathbf{x}} ,\ldots, y_{i^k_\mathbf{x}}$. 

Here $\mathbf{x}$ will be a persistence image, $y(\mathbf{x}) = 0$ will indicate the low $\beta$, confined phase, and $y(\mathbf{x}) = 1$ will indicate the high $\beta$, deconfined phase. We will train the model using data sampled from both phases close to the critical region. In the intermediate range of couplings where there is no training data, the $k$NN model will output an estimated classification $O_{k\mathrm{NN}} \in \{0,1\}$. We may then treat $\langle O_{k\mathrm{NN}} \rangle$ as a phase indicator.

As in \cite{sale2022quant}, we note that training the classifier directly on raw configurations is not computationally feasible. Doing so would require a vast number of samples to sufficiently fill out the configuration space and moreover the computational cost of the classification would grow too large. The mapping from configurations to persistence images concentrates the distribution near a low-dimensional subspace, and hence $k$NN becomes effective with far fewer samples. We also note that vectorising the persistence diagrams as persistence images is not necessary for using a $k$NN classifier since this only requires a notion of distance between samples, of which we have several for persistence diagrams. However these distances \--- such as the bottleneck distance or Wasserstein distance \--- are computationally expensive, especially for the large persistence diagrams we obtain in this application. Using the Euclidean distance between persistence images vastly speeds up the time taken to evaluate a $k$NN model and, as we will see later, maintains sufficient information to capture the phase transition.

\subsection{Finite-size Scaling Analysis}
\label{sec:FSS}

The deconfinement phase transition in the $\mathrm{SU}(2)$ lattice gauge theory is known to be a second-order transition (again thanks to the Svetitsky–Yaffe conjecture \cite{SVETITSKY1982423}). As such, quantities such as the Polyakov loop susceptibility
$$\chi(\beta) = \langle P^2 \rangle_\beta - \langle P \rangle^2_\beta$$
diverge at the phase transition in a predictable way as we approach the continuum limit. On a finite lattice this susceptibility will remain analytic, displaying a pronounced peak at a pseudo-critical $\beta$ somewhere above or below the true critical inverse coupling $\beta_c$. Holding $N_t$ (and therefore $\beta_c$) fixed, letting $N_s \rightarrow \infty$ causes the peak to narrow and move closer towards $\beta_c$. The way in which the susceptibility scales close to $\beta_c$ as a function of $N_s$ asymptotically approaches the form
\begin{equation}
\label{eqn:susceptibility_fss_secondorder}
    \chi(N_s, b) = N_s^{\gamma / \nu} \, \hat{\chi}(N_s^{1 / \nu} \, b)
\end{equation}
where $\hat{\chi}$ is a dimensionless function, $b = \frac{\beta - \beta_c}{\beta_c}$ is the reduced inverse coupling, and $\gamma$ and $\nu$ are the critical exponents for the susceptibility and correlation length respectively. By simulating close to the phase transition on different lattice sizes $N_s$ we can extract the heights and locations of the different peaks then fit these to Eq. \ref{eqn:susceptibility_fss_secondorder} to estimate $\beta_c$, $\gamma$ and $\nu$.

Similarly, persistent homology-derived observables may exhibit large variations at criticality. In Section \ref{sec:DetectTwist} we look at the difference in expectations of the birth time of a specific point in the persistence diagram under the normal action and the twisted action. In Section \ref{sec:Deconfinement} we look at the fluctuations in the output $O_{k\mathrm{NN}}$ of a trained $k$NN model, measuring the variance
\begin{align}
\begin{split}
        \chi_{k\mathrm{NN}}(\beta) & = \langle O_{k\mathrm{NN}}^2 \rangle_\beta - \langle O_{k\mathrm{NN}} \rangle^2_\beta\\ & = \langle O_{k\mathrm{NN}} \,\, \rangle_\beta (1 - \langle O_{k\mathrm{NN}} \rangle_\beta).
\end{split}
\end{align}
Note that the second equation follows since $O_{k\mathrm{NN}}$ takes values in $\{0, 1\}$. We find evidence that both these quantities display finite-size scaling behaviour similar to Eq. \ref{eqn:susceptibility_fss_secondorder} which we will use to estimate the critical inverse coupling $\beta_c$ and the critical exponent of correlation length $\nu$ via a curve collapse approach, plotting the observable against $x = N_s^{1/\nu} \, b$ for multiple lattice sizes simultaneously and finding values of $\nu$ and $\beta_c$ that minimise the distance between the curves using the Nelder-Mead method, as in the procedure described in \cite{Bhattacharjee2001AMO}.

If $\nu$ is known, we estimate $\beta_c$ by fitting the peak temperatures $\beta_c(N_s)$ of $\chi_{k\mathrm{NN}}$ obtained from multiple lattice sizes to the ansatz
\begin{equation}
\label{eqn:2nd_order_tc_scaling}
    \beta_c(N_s) - \beta_c(\infty) \propto \frac{1}{N_s^{1/\nu}}.
\end{equation}

\section{Analysis}
\label{sec:Analysis}

\subsection{Detecting Twisted Boundary Conditions}
\label{sec:DetectTwist}

We first investigate the ability of the persistent homology of our filtered complex to identify an inserted thin vortex, obtained using twisted boundary conditions, as a function of $\beta$. For $N_s \in \{ 12, 16, 20 \}$, fixing $N_t = 4$, we generate $200$ configurations using the Wilson action (\ref{eqn:action}) and $200$ configurations using the twisted action (\ref{eqn:tbc_action}) for each $\beta \in \{ 1.5, 1.6, \dots 2.9 \}$. Configurations are generated using the HiRep software \cite{PhysRevD.81.094503} with 1 heatbath step and 4 overelaxation steps for each Monte Carlo step and a sample taken every 100 Monte Carlo steps.

\begin{figure}[ht]
    \centering
    \scalebox{0.38}{\includegraphics{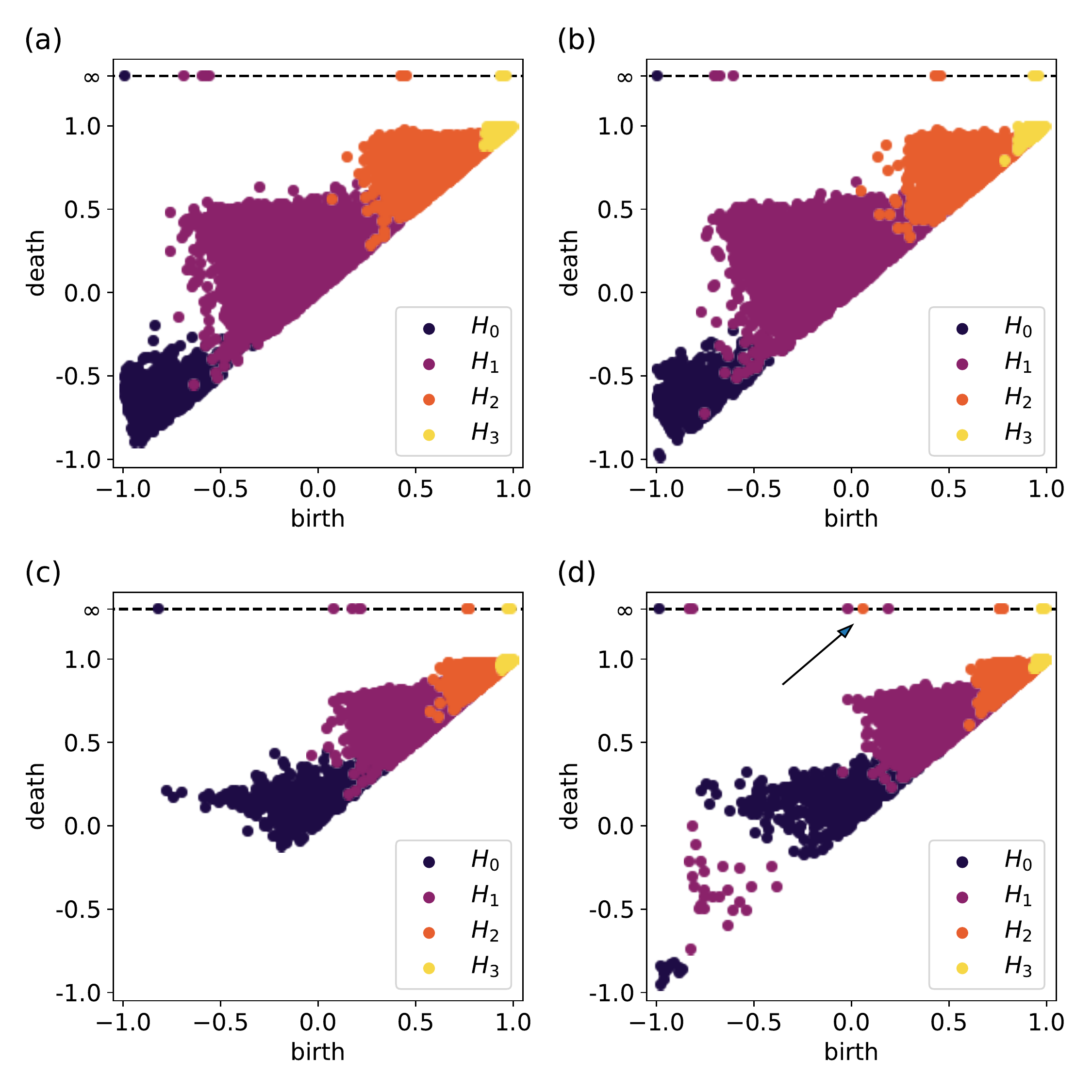}}
    \caption{Sample persistence diagrams of individual configurations obtained using the following actions and values of $\beta$: (a) Wilson, $\beta = 1.5$ (b) twisted, $\beta = 1.5$ (c) Wilson, $\beta = 2.9$ (d) twisted, $\beta = 2.9$. The arrow in (d) indicates the point $(b, \infty) \in PH_2$ with the smallest birth index $b$. Note the distance between it and the others.}
    \label{fig:PDs}
\end{figure}

Since the inserted vortex forms a closed surface, we expect to observe it in the $PH_2$ diagram. Moreover, the surface wraps round the periodic boundary of the lattice in the latter two spatial directions. If it is the first such surface to wrap around those dimensions to enter the filtered complex, then we will observe it as a point in $PH_2$ with infinite death index since it encloses a two-dimensional hole which remains even in the final complex of the filtered complex (homeomorphic to a 4-torus). Otherwise, it would appear as a point with finite death index. In Figure \ref{fig:PDs} we compare the persistence diagrams of individual sampled configurations in each phase using the Wilson action and the twisted action. In the confined phase there is no immediate distinction to be made between the persistence diagrams generated using the different actions. We claim that this is because vortices in this phase percolate throughout the system so there are likely to be many vortices that wrap around the periodic boundary conditions of the lattice. Our inserted vortex may then appear as a single point of finite persistence in $PH_2$, but the persistence of the system is not affected largely. However in the deconfined phase phase there is a clear difference. There is unlikely to be any system-generated vortex surfaces that wrap around the lattice so the inserted vortex becomes the first such surface to enter the filtered complex. We therefore observe that one of the $PH_2$ points with infinite death time has much lower birth time than the others, allowing us to identify that this point represents our inserted vortex surface. We provide additional evidence for this claim in Appendix \ref{appendix:m2_dist} by showing that the 2-cycle responsible for this point spans the same plane as the surface along which the boundary conditions are twisted. Moreover, we observe a significant change in $PH_0$ and $PH_1$ with many low persistence points appearing early on in the filtered complex. These arise as different plaquettes of the inserted vortex enter the filtered complex at different indices, forming transient connected components and holes. This is supported by the fact that these points all die by the time we reach the birth index of the point in $PH_2$.

\begin{figure}[ht]
    \centering
    \scalebox{0.56}{\includegraphics{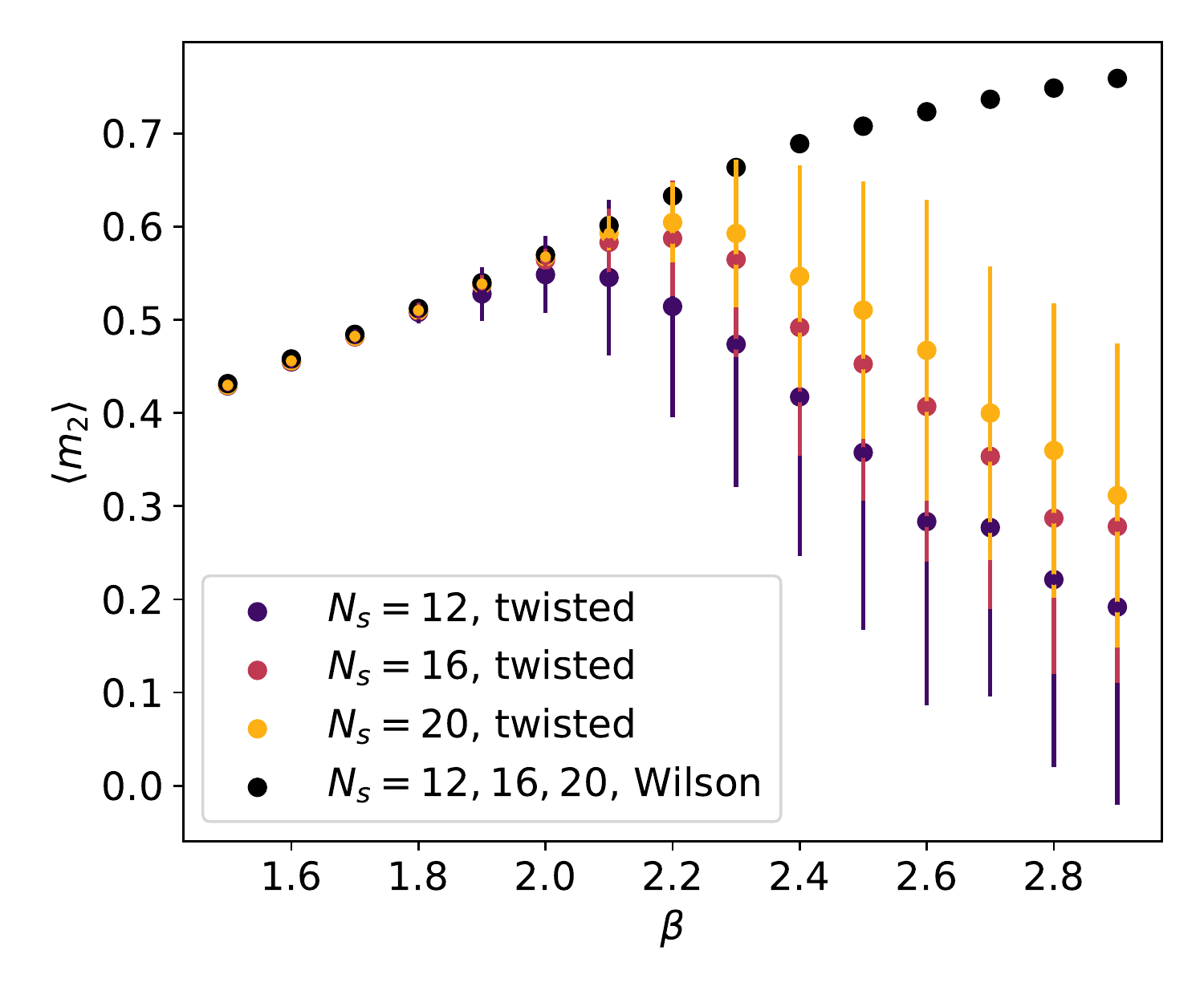}}
    \caption{The expectation value of the observable $m_2$ as a function of $\beta$ plotted for different values of $N_s$ and with the Wilson and twisted actions. The difference between the values for different lattice sizes using the Wilson action is not distinguishable at this scale, so they are plotted as the same points. Their error bars are also small enough so as to not be visible.}
    \label{fig:m2}
\end{figure}

\begin{figure}[ht]
    \centering
    \scalebox{0.56}{\includegraphics{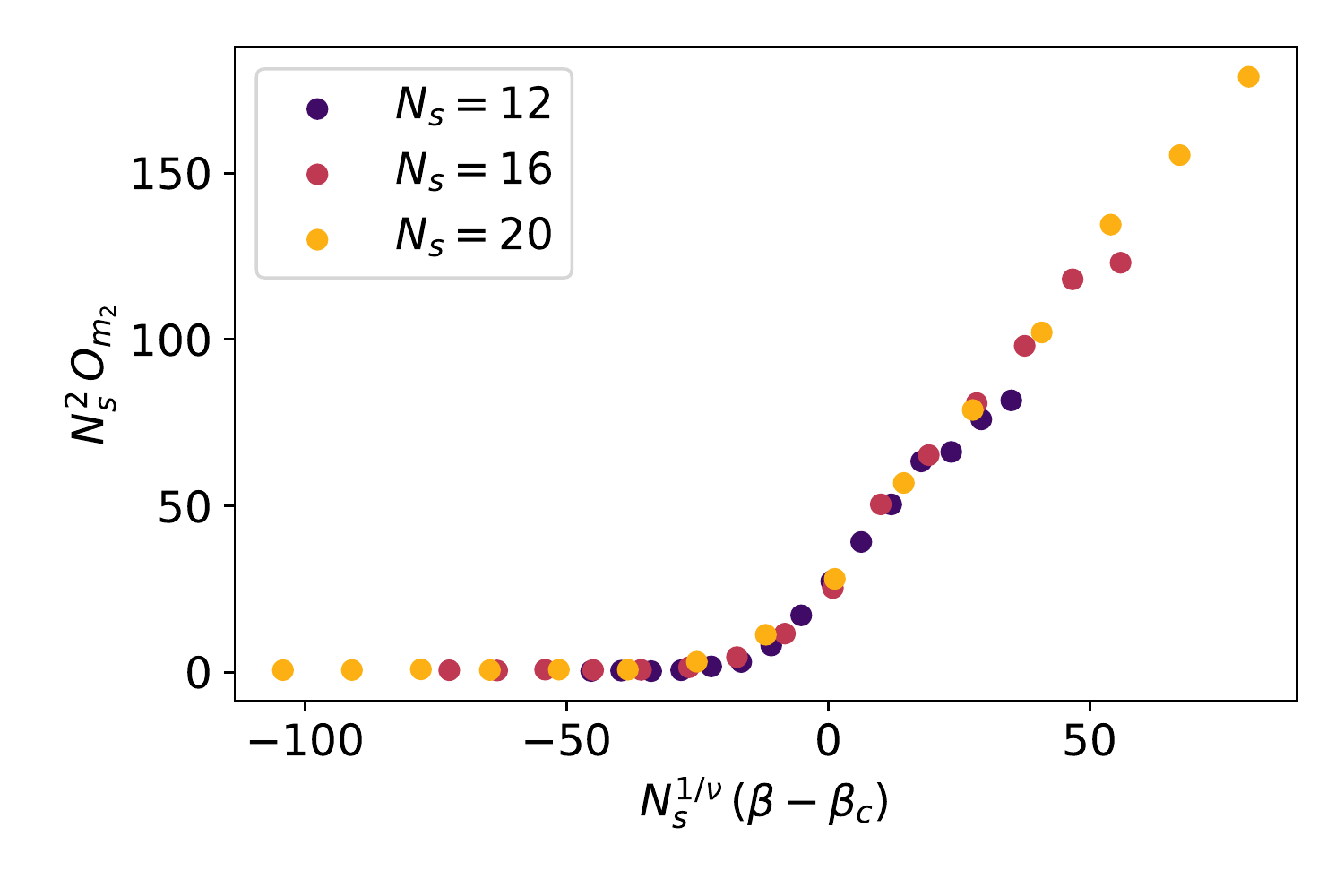}}
    \caption{The curve collapse of our phase indicator $O_{m_2}$ using $\beta_c = \twistB$ and $\nu = \twistN$. Error bars are not shown for clarity but are comparable to those in Figure \ref{fig:m2}.}
    \label{fig:O_m2}
\end{figure}

Following the discussion above, we define an observable based on the persistence diagram of a configuration:
$$m_2 = \min \big\{ \, b \,\, \big\vert \,\, (b,\infty) \in PH_2 \, \big\}.$$
The expectation value of $m_2$ for different lattice sizes with the Wilson action and twisted action are shown in Figure \ref{fig:m2}. Note that there is no difference between the expectations estimated using the different actions well into the confined phase (low $\beta$), but in the deconfined phase (high $\beta$) the curves split apart. As the lattice size increases, the point at which the curves diverge approaches the critical $\beta$ of the phase transition from below. These observations motivate measuring the difference between the expectation values using different actions
$$O_{m_2} = \langle m_2 \rangle - \langle m_2 \rangle_{\text{twist}}$$
as a phase indicator which will be zero in the confined phase and non-zero in the deconfined phase, similar to the definition of an order parameter but without the requirement to detect any symmetry breaking. A finite-size scaling analysis of this quantity yields the curve collapse in Figure \ref{fig:O_m2} with estimates of $\beta_c$ and $\nu$
\begin{align*}
    \beta_c &= \twistB \pm \twistBE\\
    \nu &= \twistN \pm \twistNE
\end{align*}
in agreement with the existing estimate $\beta_c = 2.2986(6)$ in Table \ref{tab:crit_betas}. Error estimates are obtained by performing $2000$ bootstraps. While the error obtained is reasonably large, it should be stressed that these estimates were obtained using only $200$ configurations. In Appendix \ref{appendix:m2_dist} we look more closely at the distribution of $m_2$ as measured using the twisted action in order to better understand the behaviour of $O_{m_2}$.

Note that we fixed the exponent of $-2$ for the scaling of $O_{m_2}$ with $N_s$. Attempting to fit this exponent along with $\beta_c$ and $\nu$ often led to the optimizer returning unrealistic large positive values for the exponent, spoiling the error estimation. The value of $-2$ was found by hand to give a good fit and we offer a heuristic argument for why. First note that $N_s^2$ is how the number of plaquettes in the inserted vortex surface scales with $N_s$, since the surface wraps round the periodic boundary conditions. In the case where it describes the formation of the inserted vortex surface, the value of $m_2$ is determined by the filtration index of the last plaquette to enter the surface. Now the larger the surface, the more likely it is that there will be at least one plaquette in the surface affected by noise or pierced by another vortex, causing it to enter the filtered complex late and dragging the value of $m_2$ closer to its average in the Wilson action. Assuming that this likelihood is independent for each plaquette or at least approximately linear in the number of plaquettes, we therefore obtain the quadratic scaling in $N_s$.

\subsection{Investigating Deconfinement Without Twisted Boundary Conditions}
\label{sec:Deconfinement}

We now investigate whether the persistent homology is able to detect the deconfinement phase transition purely from the Wilson action by making use of a simple machine learning framework inspired by that in \cite{sale2022quant}. In particular, for $N_t \in \{ 4, 5, 6 \}$ we attempt to estimate the critical inverse coupling $\beta_c$ and the critical exponent of the correlation length $\nu$ via a finite-size scaling analysis of the output of a $k$-nearest neighbours classifier trained on the persistence. In each case we repeat the following procedure for each $N_s \in \{12, 16, 20, 24 \}$.
\begin{enumerate}
    \item Configurations are sampled from a range of values of $\beta$ (specific values are given in each case below) using the HiRep software \cite{PhysRevD.81.094503} with 1 heatbath step and 4 overelaxation steps for each Monte Carlo step and a sample taken every 100 Monte Carlo steps.
    \item For each sample we compute their $PH_0$, $PH_1$, $PH_2$ and $PH_3$ persistent homology using our filtered complex and compute the corresponding persistence images with a resolution of $25 \times 25$ and $\sigma$ equal to $5\%$ of a pixel. We concatenate the 4 separate images and flatten them into a $4 \times 25 \times 25 = 2500$ dimensional vector.
    \item We train a $k$NN model ($k=30$) to predict the phase of a configuration based on its concatenated persistence image vector by using vectors from well into the confined and deconfined phases.
    \item Using the trained classification model, we define an observable $O_{k\mathrm{NN}}$ which is the predicted phase of a given configuration.
    \item Applying multiple histogram reweighting to the variance
    \begin{align*}
        \chi_{k\mathrm{NN}} &= \langle O_{k\mathrm{NN}}^2 \rangle - \langle O_{k\mathrm{NN}} \rangle^2\\
        &= \langle O_{k\mathrm{NN}} \rangle \, (1 - \langle O_{k\mathrm{NN}} \rangle)
    \end{align*}
     (where the second equality follows since $O_{k\mathrm{NN}} \in \{0,1\}$), we obtain an interpolated curve and a more precise estimate of the location of its peak.
\end{enumerate}
By performing a finite-size scaling analysis of the locations of the peaks obtained for each value of $N_s$, and a curve collapse of the different reweighted variance curves, we obtain estimates of $\beta_c$ and $\nu$ for the deconfinement phase transition at the given value of $N_t$.

\subsubsection{$N_t = 4$}
\label{sec:Nt4}

For lattices of size $4 \times N_s^3$ with $N_s \in \{ 12, 16, 20, 24 \}$, we train a $k$-nearest neighbours classifier ($k = 30$) on the concatenated $PH_0$, $PH_1$, $PH_2$ and $PH_3$ persistence images of $200$ configurations sampled at each $\beta$ in the confined and deconfined regions given in Table \ref{tab:Nt4_betas}. The classifier is then used to produce a predicted classification $O_{k\mathrm{NN}}$ for $200$ configurations sampled for each value of $\beta$ from the critical region. 

\begin{table}[h!]
\centering
\begin{tabular}{ || P{5em} | P{19em} || }
 \hline
 Region & $\beta$ \\
 \hline
 Confined & 2.2 , 2.21, 2.22, 2.23, 2.24 \\ \hline

 Deconfined & 2.36, 2.37, 2.38, 2.39, 2.4 \\ \hline

 Critical & 2.25, 2.26, 2.27, 2.275, 2.28, 2.285, 2.29, 2.295, 2.298, 2.299, 2.3, 2.301, 2.302, 2.305, 2.31, 2.315, 2.32, 2.325, 2.33, 2.34, 2.35 \\
 \hline
\end{tabular}
\caption{Values of $\beta$ sampled at for the $N_t = 4$ phase transition.}
\label{tab:Nt4_betas}
\end{table}

\begin{figure}[ht]
    \centering
    \scalebox{0.56}{\includegraphics{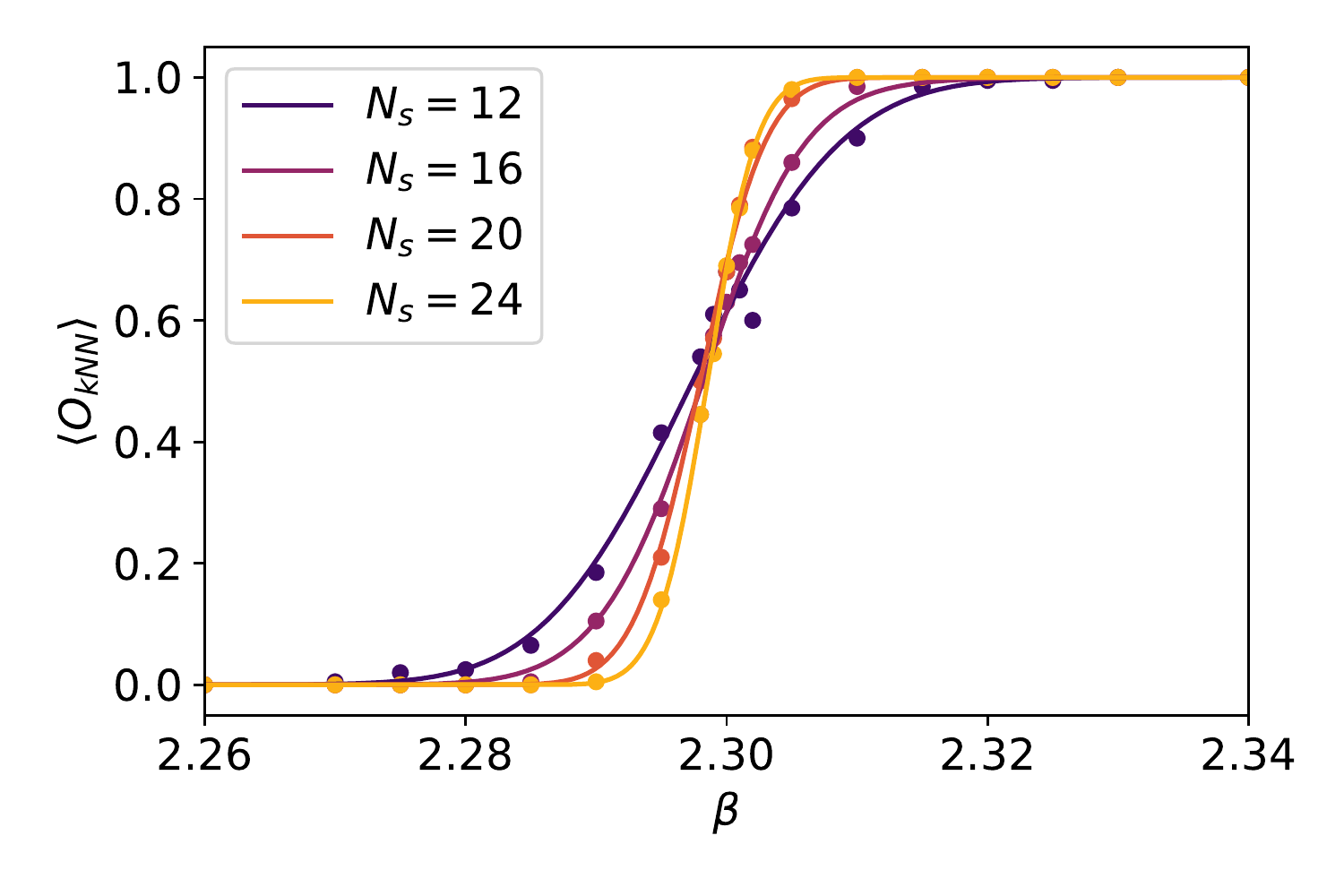}}
    \caption{Plot showing our phase indicator $\langle O_{k\mathrm{NN}} \rangle$ as a function of $\beta$ for $N_t = 4$. The points show the measured expectations and the curve is the output of histogram reweighting these measurements.}
    \label{fig:Oknn}
\end{figure}

\begin{figure}[ht]
    \centering
    \scalebox{0.56}{\includegraphics{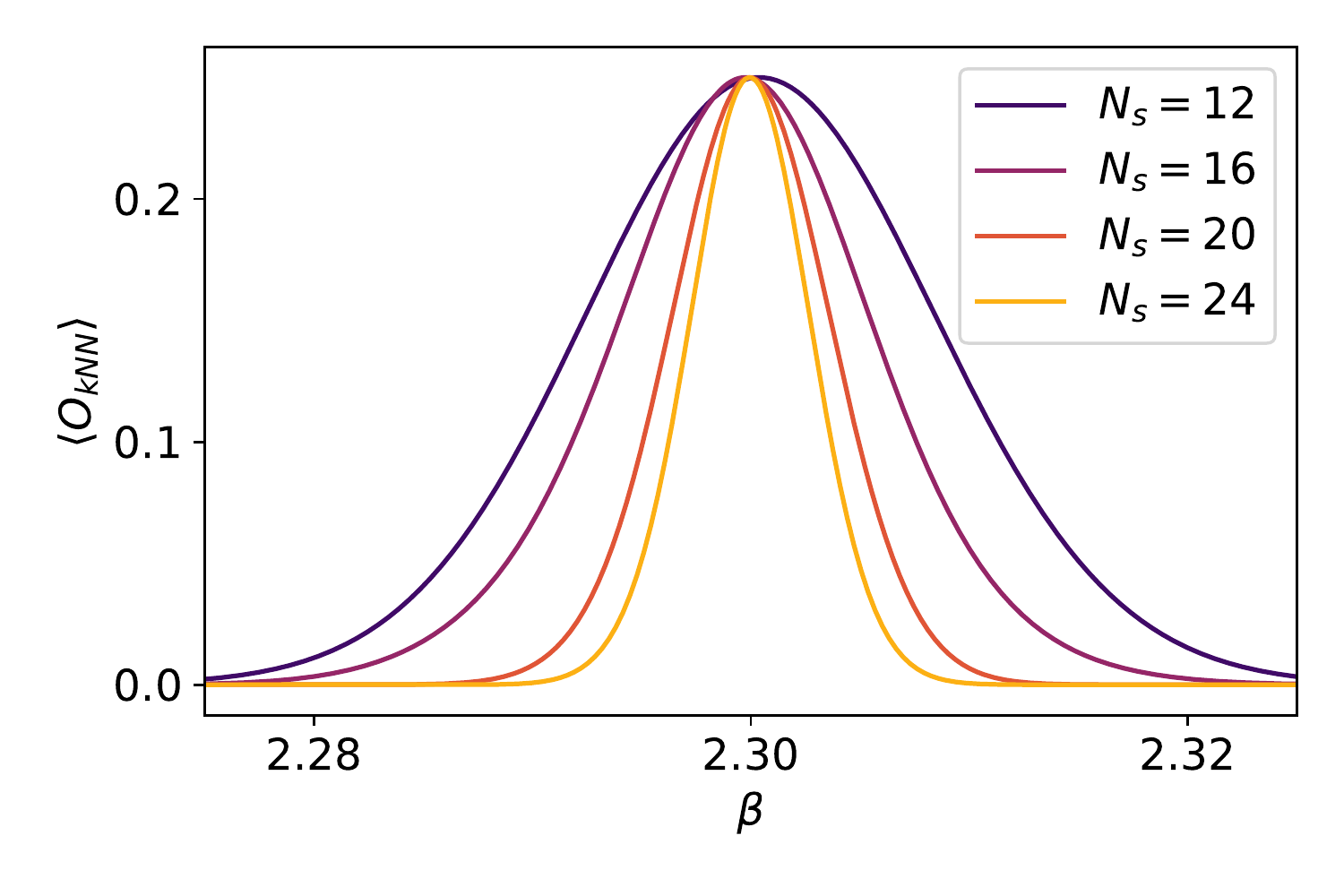}}
    \caption{The variance curves $\chi_{k\mathrm{NN}}$ of $O_{k\mathrm{NN}}$ for $N_t = 4$ to which we will apply our curve collapse procedure.}
    \label{fig:Xknn}
\end{figure}

The resulting estimates of the expectation $\langle O_{k\mathrm{NN}} \rangle(\beta)$ are shown in Figure \ref{fig:Oknn} along with interpolating curves obtained via histogram reweighting. The variance curves $\chi_{k\mathrm{NN}}$ are shown in Figure \ref{fig:Xknn}.

From here we proceed with two separate analyses. In the first we estimate $\beta_c$ from a linear regression assuming the value of the critical exponent $\nu$ to be known. In the second we estimate $\beta_c$ and $\nu$ concurrently via a curve collapse procedure. In both cases we perform two separate bootstraps to obtain $500$ bootstrap samples from each. One bootstrap is carried out by resampling the configurations for each $\beta$ used to train the $k$-nearest neighbours classifier. The other is carried out by resampling the configurations for each $\beta$ used to estimate $\langle O_{k\mathrm{NN}} \rangle$. Applying the finite-size scaling analysis to both collections of bootstraps yields two separate distributions for $\beta_c$ and two for $\nu$. The error in these quantities is therefore estimated by combining the standard deviation of the distributions coming from the different bootstrap procedures under the assumption that they are independent.

\begin{figure}[ht]
    \centering
    \scalebox{0.56}{\includegraphics{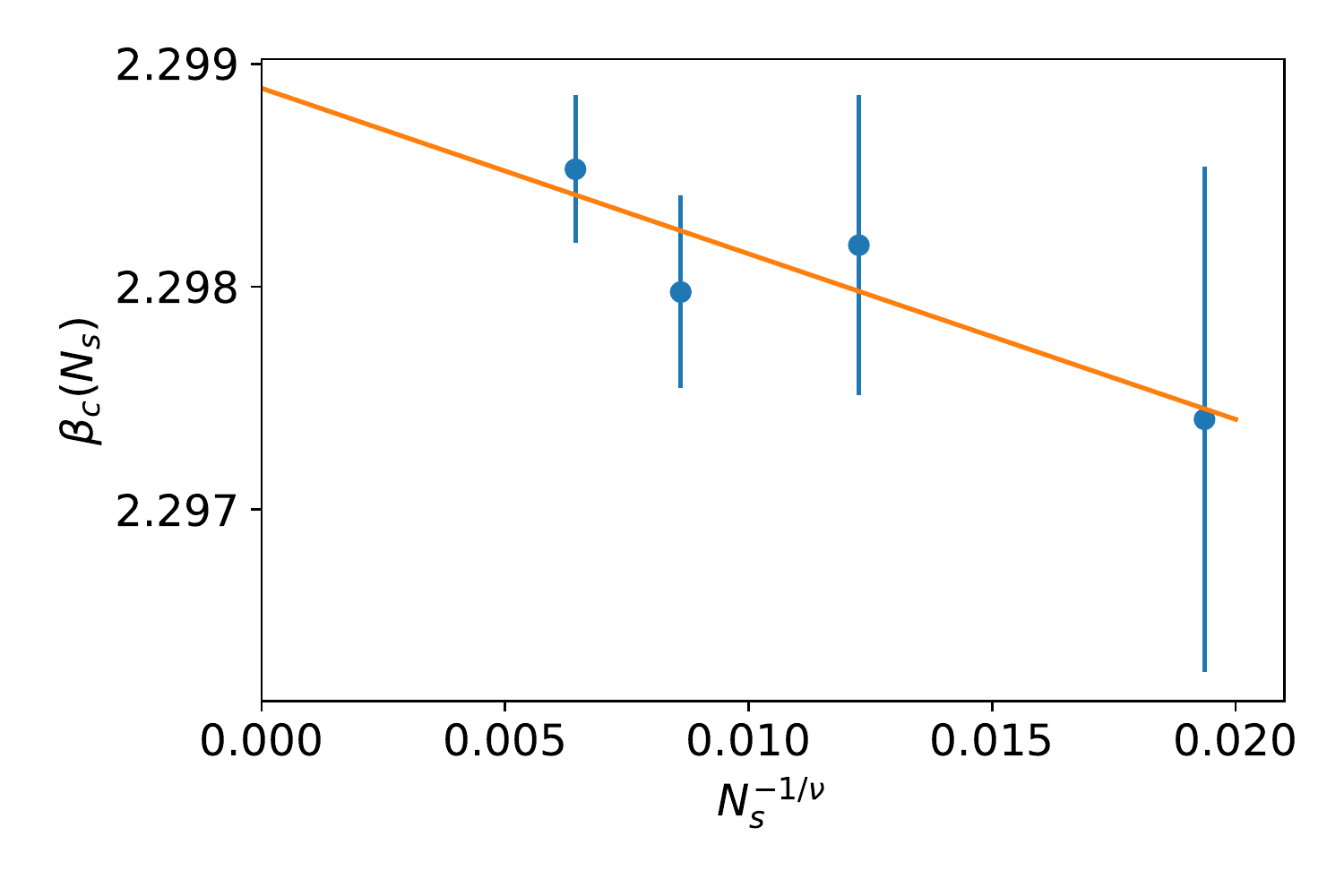}}
    \caption{Estimating $\beta_c$ for $N_t = 4$. The pseudo-critical values of $\beta$, obtained from locating the peaks of the variance curves in Figure \ref{fig:Xknn}, are fitted to the ansatz (\ref{eqn:2nd_order_tc_scaling}). Error bars are estimated by bootstrapping.}
    \label{fig:bc_line_Nt4}
\end{figure}

By defining the pseudo-critical inverse coupling $\beta_c(N_s)$ to be the point at which $\chi_{k\mathrm{NN}}$ peaks, we can plot $\beta_c(N_s)$ against $N_s^{-1/\nu}$ using a previously estimated \cite{precision_ising} value of $\nu = 0.629971$. The result is shown in Figure \ref{fig:bc_line_Nt4}. We observe that the points plotted with error bars of $1 \sigma$ support a straight line fit. The intercept yields $\beta_c = \linknnBfour \pm \linknnBEfour$, supporting the previously obtained estimate of $\beta_c = 2.2986(6)$ in Table \ref{tab:crit_betas}.

To estimate $\beta_c$ and $\nu$ concurrently we employ a numerical curve collapse procedure, plotting $\chi_{k\mathrm{NN}}$ against $N_s^{1/\nu} \, (\beta - \beta_c)$ and tuning $\beta_c$ and $\nu$ to minimise the distance between the curves using the Nelder-Mead method.

\begin{figure}[ht]
    \centering
    \scalebox{0.56}{\includegraphics{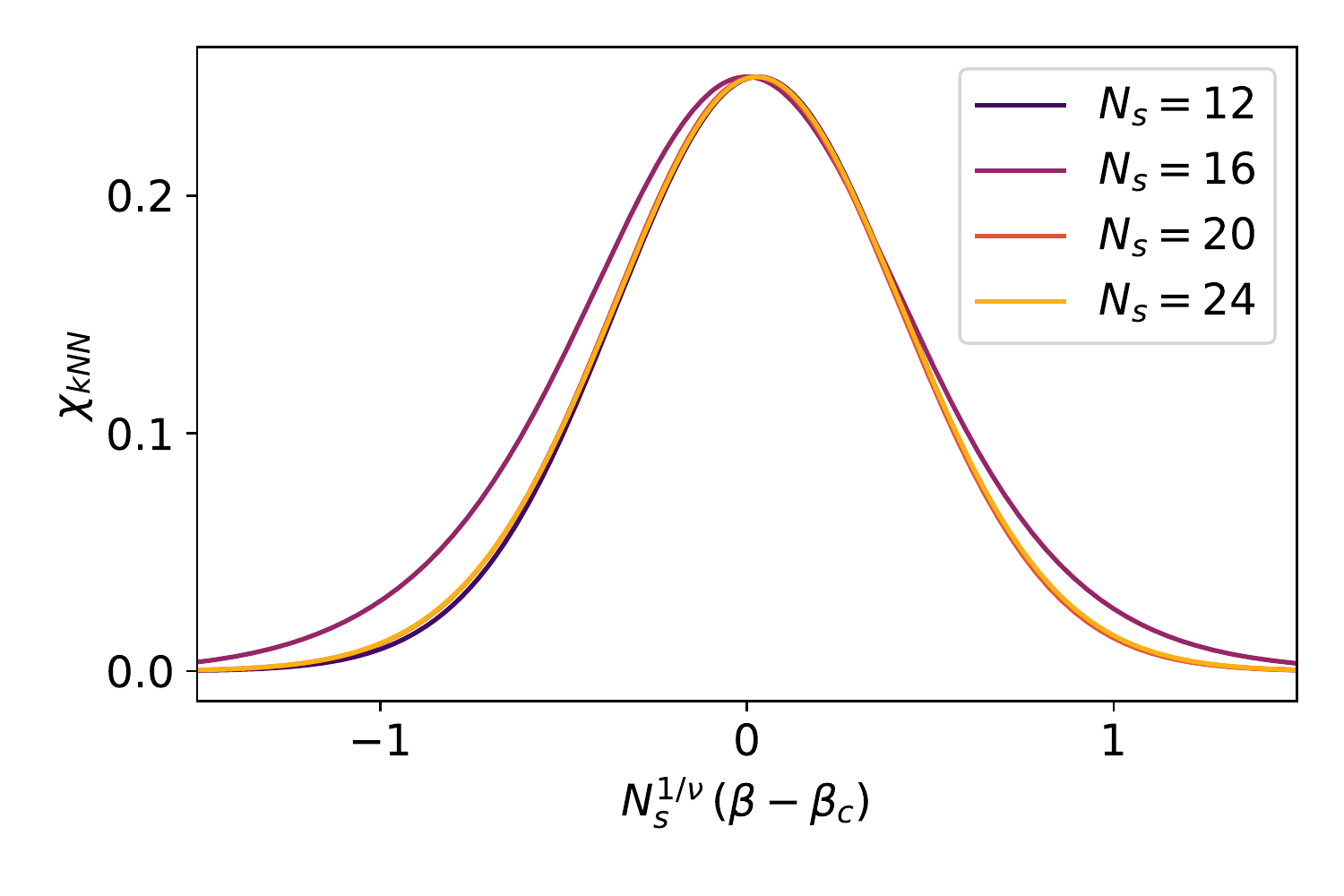}}
    \caption{The curve collapse of $\chi_{k\mathrm{NN}}$ for $N_t = 4$ using $\beta_c = \knnBfour$ and $\nu = \knnNfour$.}
    \label{fig:collapsed_Xknn_Nt4}
\end{figure}

The resulting curve collapse is shown in Figure \ref{fig:collapsed_Xknn_Nt4} and the obtained estimates of $\beta_c$ and $\nu$
\begin{align*}
    \beta_c &= \knnBfour \pm \knnBEfour\\
    \nu &= \knnNfour \pm \knnNEfour
\end{align*}
are consistent with previous estimates.

To confidently claim that this methodology identifies the phase transition, we also tried using alternative values of $\beta$ to train the $k$NN classifier, chosen further away from the transition point and so that the transition point is further from the center point between the highest $\beta$ in the confined phase and the lowest $\beta$ in the deconfined phase. The alternative training values are shown in Table \ref{tab:Nt4_betas_alt}.

\begin{table}[h!]
\centering
\begin{tabular}{ || P{5em} | P{19em} || }
 \hline
 Region & $\beta$ \\
 \hline
 Confined & 2.21, 2.22, 2.23 \\ \hline

 Deconfined & 2.38, 2.39, 2.4 \\ \hline

 Critical & 2.24, 2.25, 2.26, 2.27, 2.275, 2.28, 2.285, 2.29, 2.295, 2.298, 2.299, 2.3, 2.301, 2.302, 2.305, 2.31, 2.315, 2.32, 2.325, 2.33, 2.34, 2.35, 2.36, 2.37 \\
 \hline
\end{tabular}
\caption{Alternative values of $\beta$ sampled at for the $N_t = 4$ phase transition to test the sensitivity of the method to the choice of training data.}
\label{tab:Nt4_betas_alt}
\end{table}

Using these training values we obtain estimates from the linear fit of
\begin{align*}
    \beta_c &= \linknnBfouralt \pm \linknnBEfouralt
\end{align*}
and the curve collapse of
\begin{align*}
    \beta_c &= \knnBfouralt \pm \knnBEfouralt\\
    \nu &= \knnNfouralt \pm \knnNEfouralt,
\end{align*}
close to our previous ones and still compatible with our reference estimates.

\subsubsection{$N_t = 5$}
\label{sec:Nt5}

For lattices of size $5 \times N_s^3$ with $N_s \in \{ 12, 16, 20, 24 \}$, we train a $k$-nearest neighbours classifier ($k = 30$) on the concatenated $PH_0$, $PH_1$, $PH_2$ and $PH_3$ persistence images of $200$ configurations sampled at each $\beta$ in the confined and deconfined regions given in Table \ref{tab:Nt5_betas}. The classifier is then used to produce a predicted classification $O_{k\mathrm{NN}}$ for $200$ configurations sampled for each value of $\beta$ from the critical region. 

\begin{table}[h!]
\centering
\begin{tabular}{ || P{5em} | P{19em} || }
 \hline
 Region & $\beta$ \\
 \hline
 Confined & 2.29, 2.3, 2.31, 2.32, 2.33 \\ \hline

 Deconfined & 2.41, 2.42, 2.43, 2.44, 2.45 \\ \hline

 Critical & 2.34, 2.345, 2.35, 2.355, 2.36, 2.365, 2.369, 2.37, 2.371, 2.372, 2.375, 2.38, 2.385, 2.39, 2.395, 2.4 \\
 \hline
\end{tabular}
\caption{Values of $\beta$ sampled at for the $N_t = 5$ phase transition.}
\label{tab:Nt5_betas}
\end{table}

\begin{figure}[ht]
    \centering
    \scalebox{0.56}{\includegraphics{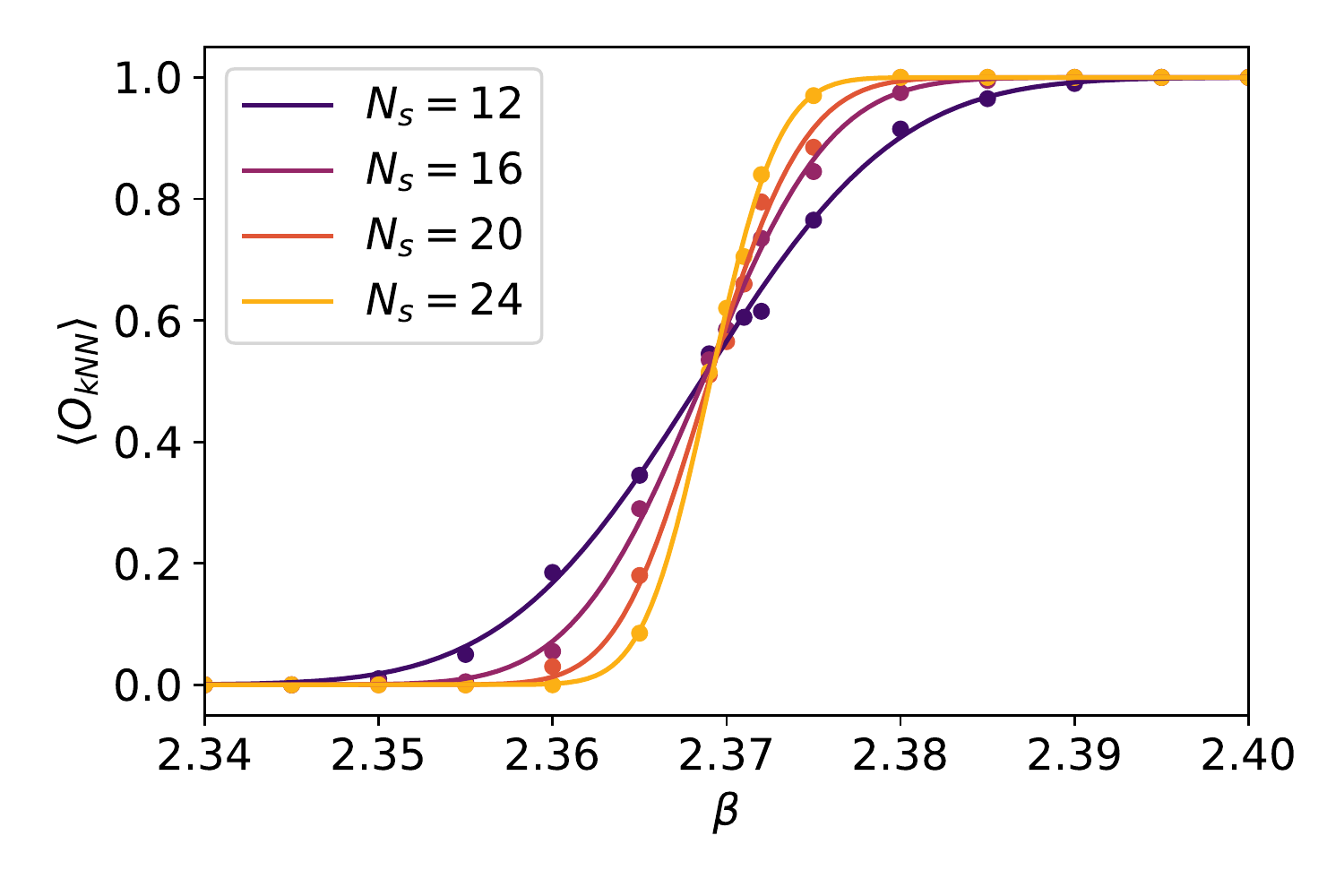}}
    \caption{Plot showing our phase indicator $\langle O_{k\mathrm{NN}} \rangle$ as a function of $\beta$ for $N_t = 5$. The points show the measured expectations and the curve is the output of histogram reweighting these measurements.}
    \label{fig:Oknn_Nt5}
\end{figure}

The resulting estimates of the expectation $\langle O_{k\mathrm{NN}} \rangle(\beta)$ are shown in Figure \ref{fig:Oknn_Nt5} along with interpolating curves obtained via histogram reweighting.

\begin{figure}[ht]
    \centering
    \scalebox{0.56}{\includegraphics{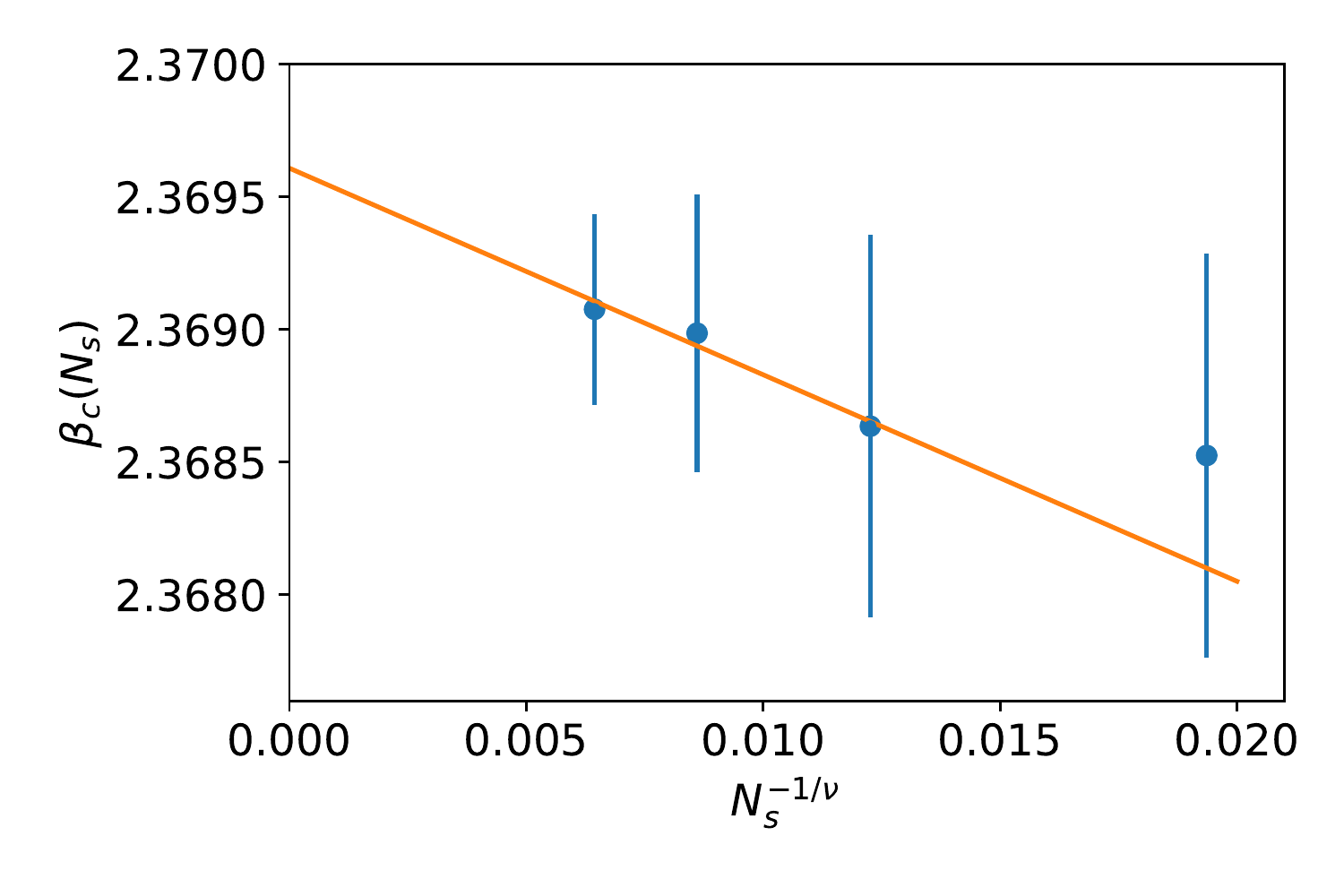}}
    \caption{Estimating $\beta_c$ for $N_t = 5$. The pseudo-critical values of $\beta$ of the largest three lattice sizes, obtained from locating the peaks of the variance curves, are fitted to the ansatz (\ref{eqn:2nd_order_tc_scaling}). Error bars are estimated by bootstrapping.}
    \label{fig:bc_line_Nt5}
\end{figure}

The plot of the pseudo-critical $\beta_c(N_s)$ against $N_s^{-1/\nu}$ using $\nu = 0.629971$ is shown in Figure \ref{fig:bc_line_Nt5}. Here we fit a straight line to the largest three lattice sizes since this gives a better fit than including $N_s = 12$. The intercept yields $\beta_c = \linknnBfive \pm \linknnBEfive$ which is just about compatible with the previously obtained estimate of $\beta_c = 2.37136(54)$ in Table \ref{tab:crit_betas}.

\begin{figure}[ht]
    \centering
    \scalebox{0.56}{\includegraphics{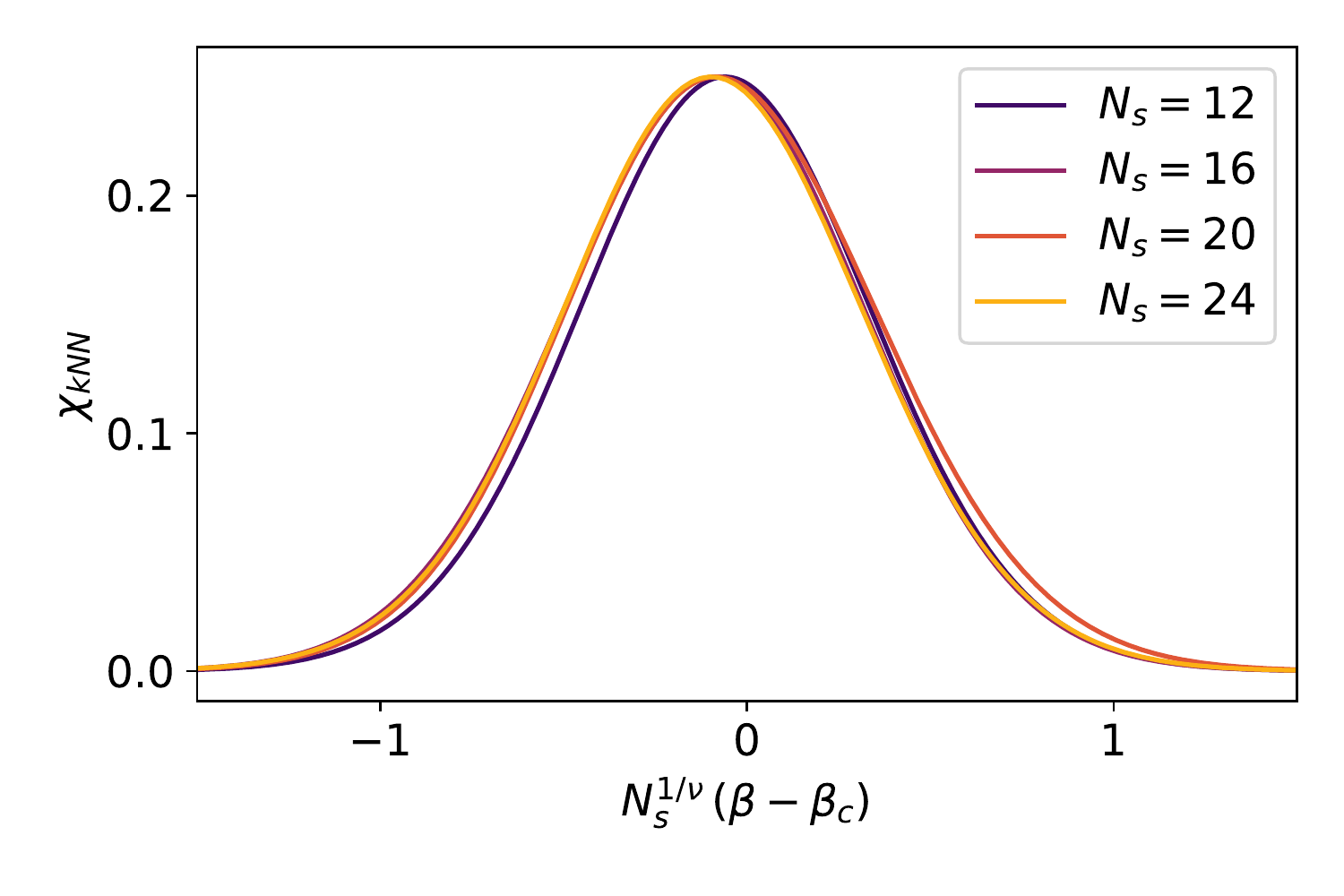}}
    \caption{The curve collapse of $\chi_{k\mathrm{NN}}$ for $N_t = 5$ using $\beta_c = \knnBfive$ and $\nu = \knnNfive$.}
    \label{fig:collapsed_Xknn_Nt5}
\end{figure}

We also perform the curve collapse on only the highest three lattice sizes and the result is shown in Figure \ref{fig:collapsed_Xknn_Nt5}. The obtained estimates of $\beta_c$ and $\nu$
\begin{align*}
    \beta_c &= \knnBfive \pm \knnBEfive \\
    \nu &= \knnNfive \pm \knnNEfive
\end{align*}
are consistent with previous estimates.

\subsubsection{$N_t = 6$}
\label{sec:Nt6}

For lattices of size $6 \times N_s^3$ with $N_s \in \{ 12, 16, 20, 24 \}$, we train a $k$-nearest neighbours classifier ($k = 30$) on the concatenated $PH_0$, $PH_1$, $PH_2$ and $PH_3$ persistence images of $200$ configurations sampled at each $\beta$ in the confined and deconfined regions given in Table \ref{tab:Nt6_betas}. The classifier is then used to produce a predicted classification $O_{k\mathrm{NN}}$ for $200$ configurations sampled for each value of $\beta$ from the critical region. 

\begin{table}[h!]
\centering
\begin{tabular}{ || P{5em} | P{19em} || }
 \hline
 Region & $\beta$ \\
 \hline
 Confined & 2.33, 2.34, 2.35, 2.36, 2.37 \\ \hline

 Deconfined & 2.49, 2.5, 2.51, 2.52, 2.53 \\ \hline

 Critical & 2.38, 2.39, 2.4, 2.405, 2.41, 2.415, 2.42, 2.425, 2.426, 2.427, 2.428, 2.43, 2.435, 2.44, 2.445, 2.45, 2.455, 2.46, 2.47, 2.48 \\
 \hline
\end{tabular}
\caption{Values of $\beta$ sampled at for the $N_t = 6$ phase transition.}
\label{tab:Nt6_betas}
\end{table}

\begin{figure}[ht]
    \centering
    \scalebox{0.56}{\includegraphics{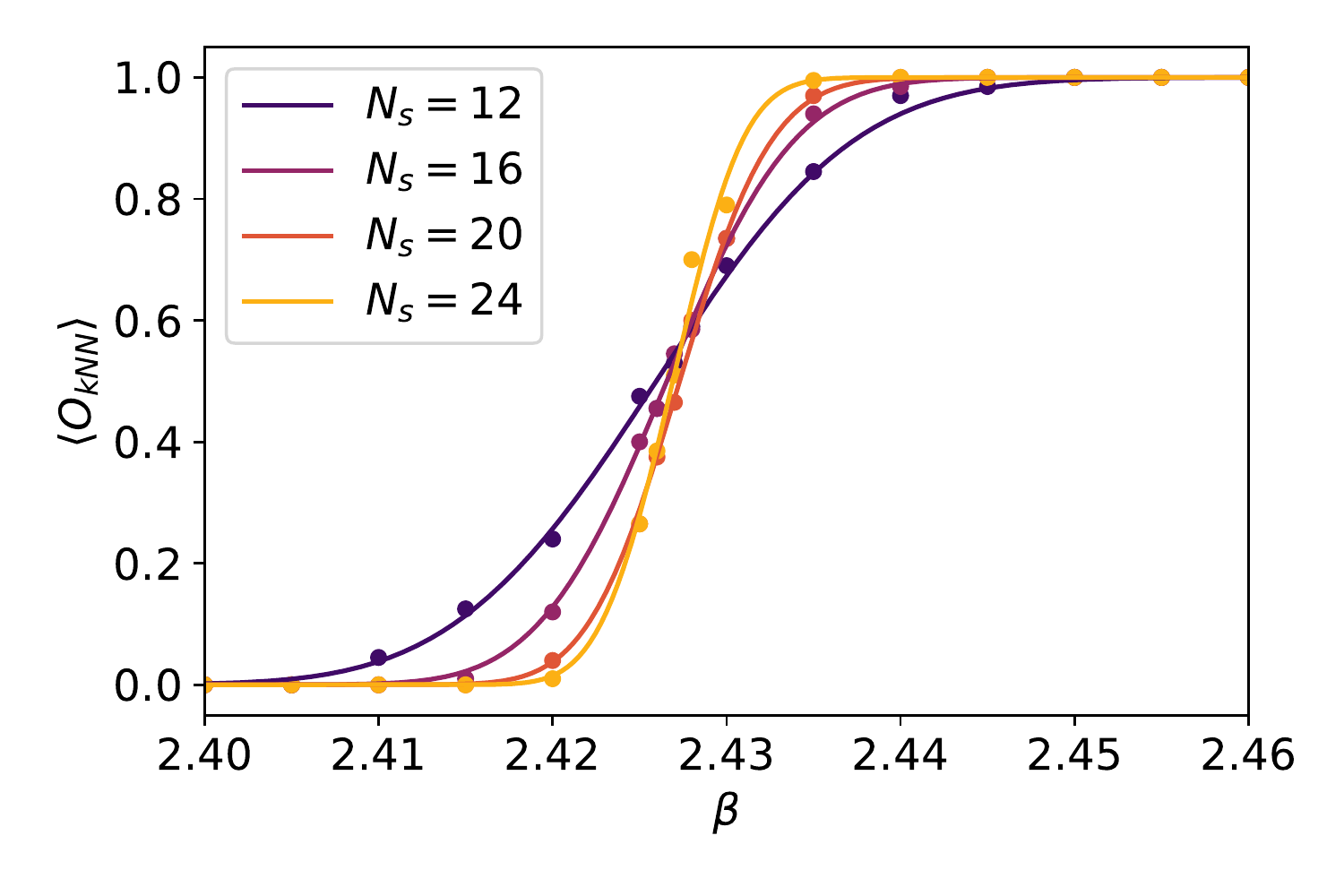}}
    \caption{Plot showing our phase indicator $\langle O_{k\mathrm{NN}} \rangle$ as a function of $\beta$ for $N_t = 6$. The points show the measured expectations and the curve is the output of histogram reweighting these measurements.}
    \label{fig:Oknn_Nt6}
\end{figure}

 The resulting estimates of the expectation $\langle O_{k\mathrm{NN}} \rangle(\beta)$ are shown in Figure \ref{fig:Oknn_Nt6} along with interpolating curves obtained via histogram reweighting.

\begin{figure}[ht]
    \centering
    \scalebox{0.56}{\includegraphics{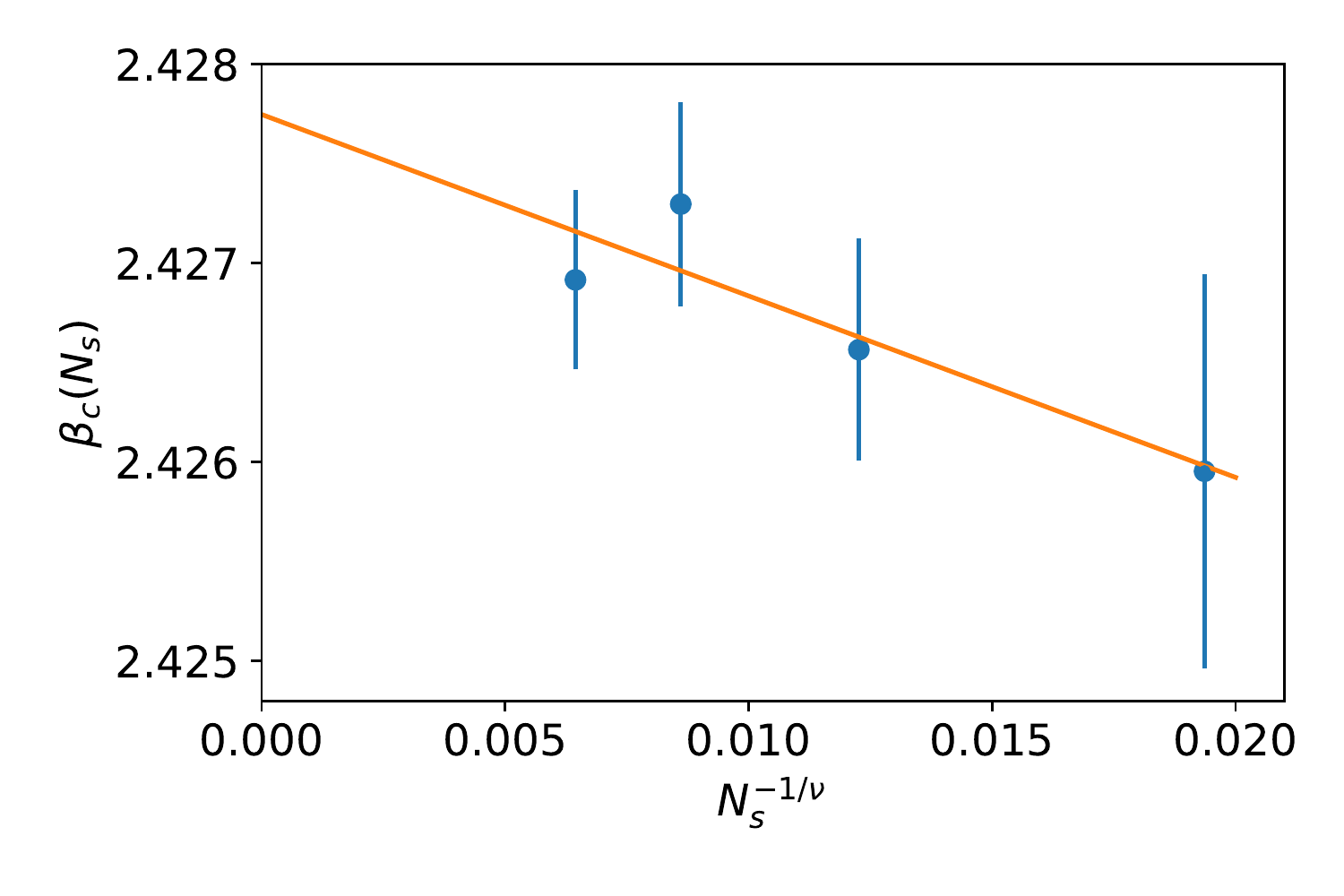}}
    \caption{Estimating $\beta_c$ for $N_t = 6$. The pseudo-critical values of $\beta$, obtained from locating the peaks of the variance curves, are fitted to the ansatz (\ref{eqn:2nd_order_tc_scaling}). Error bars are estimated by bootstrapping.}
    \label{fig:bc_line_Nt6}
\end{figure}

The plot of the pseudo-critical $\beta_c(N_s)$ against $N_s^{-1/\nu}$ using $\nu = 0.629971$ is shown in Figure \ref{fig:bc_line_Nt6}. The intercept of the straight line fit yields $\beta_c = \linknnBsix \pm \linknnBEsix$, supporting the previously obtained estimate of $\beta_c = 2.4271(17)$ in Table \ref{tab:crit_betas}.

\begin{figure}[ht]
    \centering
    \scalebox{0.56}{\includegraphics{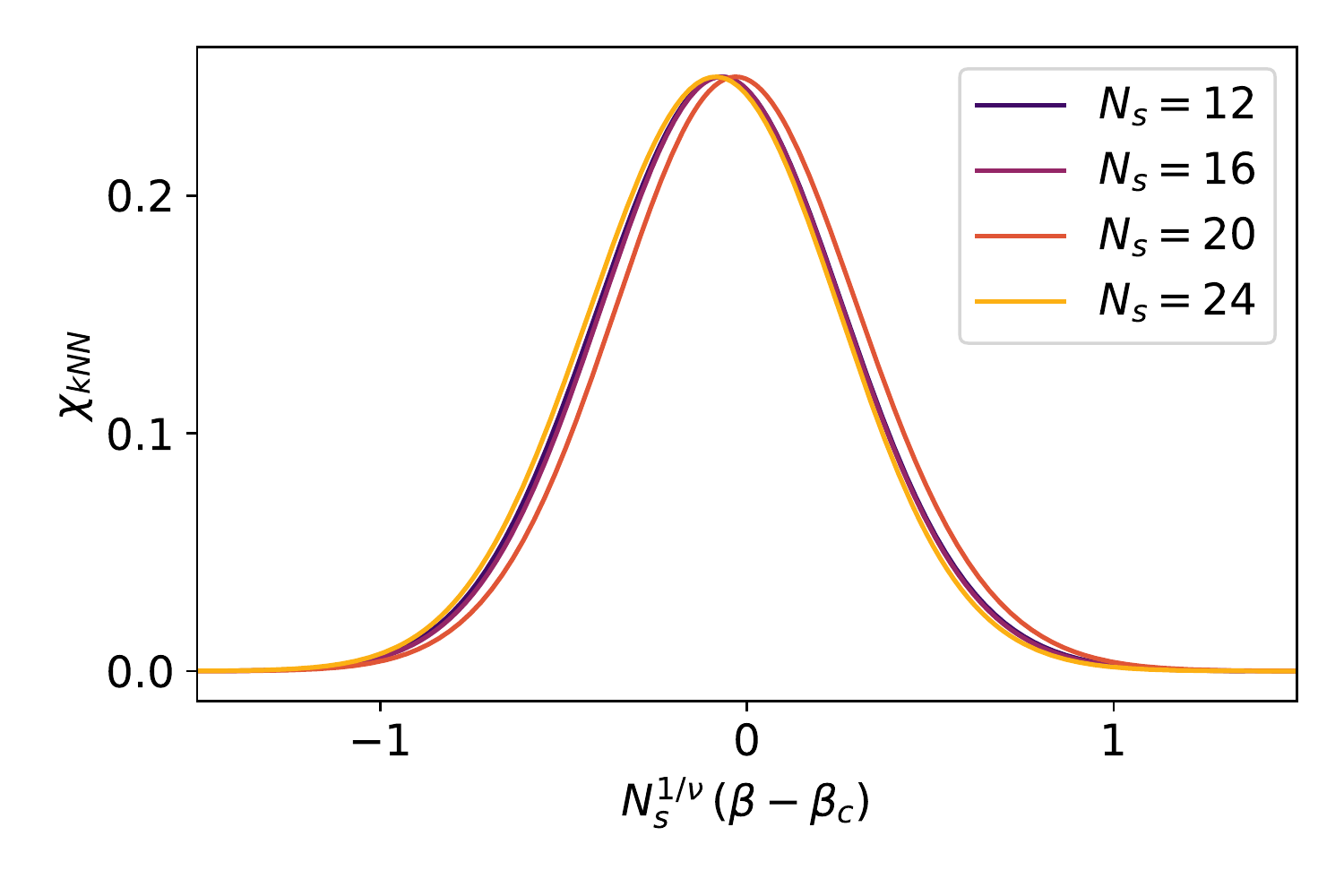}}
    \caption{The curve collapse of $\chi_{k\mathrm{NN}}$ for $N_t = 6$ using $\beta_c = \knnBsix$ and $\nu = \knnNsix$.}
    \label{fig:collapsed_Xknn_Nt6}
\end{figure}

The result of the curve collapse is shown in Figure \ref{fig:collapsed_Xknn_Nt6}. The obtained estimates of $\beta_c$ and $\nu$ are
\begin{align*}
    \beta_c &= \knnBsix \pm \knnBEsix\\
    \nu &= \knnNsix \pm \knnNEsix.
\end{align*}
The estimate of $\beta_c$ agrees with the previous estimate, however the previous estimate of $\nu = 0.629971$ we refer to lies just over 2 standard deviations outside of our estimate. We ascribe this discrepancy to the use of smaller aspect ratios $N_s/N_t$ as we increase $N_t$, remarking that the purpose of the investigation we have performed at various $N_t$ is to show reasonable scaling with the lattice spacing rather than providing a precision study of critical properties of the system. 

\section{Conclusions and Discussion}
\label{sec:Conclusions}

In this paper we have developed a gauge-invariant method using persistent homology, a tool from topological data analysis, that is designed to detect center vortices in configurations of $\mathrm{SU}(2)$ lattice gauge theory. We defined two different persistence-based phase indicators for the deconfinement phase transition --- one making use of twisted boundary conditions, and the other using simple machine learning --- and successfully used them to estimate the critical $\beta$ and exponent $\nu$ of the transition.

This method was designed to detect and capture the behaviour of vortices and we provided preliminary evidence for this by showing it is possible to detect the use of twisted boundary conditions in the deconfined phase. A closer look at the relationship between the individual points in the resulting persistence diagrams and center vortices is left to a future analysis. In particular, it would be interesting to compare representative cycles for points in the $H_2$ persistence diagram with the vortex surfaces obtained by gauge fixing and projection. Moreover we showed that it is possible to detect the deconfinement phase transition from the persistent homology of samples using the original action alone. We argue that this has the potential to lend support to the center vortex picture of confinement once the connection between the persistent homology of our filtration and center vortices is fully established, at least by not ruling it out. For a stronger argument we would need to investigate the relationship between our method and other competing pictures. For example, looking at the monopole picture of confinement (see \cite{https://doi.org/10.48550/arxiv.hep-th/0010225} or \cite{greensite2011introduction} for a summary), one could investigate the sensitivity of our method to monopoles. If the method were sensitive to vortices but not to monopoles, the fact that it captures the phase transition would be evidence for the vortex picture over the monopole picture. However, this is work yet to be done, which will require non-trivial adaptations of some of the steps used here. For instance, if the current method is indeed sensitive to monopoles, then one might instead attempt to devise a filtration that exposes monopole-like singularities but not vortices.

Besides a greater degree of interpretability, another advantage of this method over machine learning approaches based on deep learning is that we were able to obtain our results using only a small number of sampled configurations. This is particularly important in view of extending our methodology to full QCD, for which numerical computations for generating gauge configurations near the physical point are very demanding and hence the number of configurations one can use is generally limited. 

\begin{acknowledgments}
The authors would like to thank Ryan Bignell and Tin Sulejmanpasic for comments on the manuscript as well as Ed Bennett for advice and feedback on the code and data releases. Numerical simulations have been performed on the Swansea SUNBIRD system. This system is part of the Supercomputing Wales project, which is part-funded by the European Regional Development Fund (ERDF) via Welsh Government. Configurations of the $\mathrm{SU}(2)$ lattice gauge theory were sampled using the HiRep software \cite{PhysRevD.81.094503}. Persistent homology calculations were performed using giotto-tda \cite{tauzin2020giottotda}. Histogram reweighting calculations were performed using pymbar \cite{Shirts2008StatisticallyOA}. NS has been supported by a Swansea University Research Excellence Scholarship (SURES). JG was supported by EPSRC grant EP/R018472/1. BL received funding from the European Research Council (ERC) under the European Union’s Horizon 2020 research and innovation programme under grant agreement No 813942. The work of BL was further supported in part by the UKRI Science and Technology Facilities Council (STFC) Consolidated Grant ST/T000813/1, by the Royal Society Wolfson Research Merit Award WM170010 and by the Leverhulme Foundation Research Fellowship RF-2020-461{\textbackslash}9. 
\end{acknowledgments}

\appendix

\section{Cubical Complexes and Homology}
\label{appendix:cubical}

This is a very compressed version of the exposition found in \cite{kaczynski2004computational}. An \textit{elementary interval} is an interval of the form $[i, i+1] \subset \reals$ (\textit{non-degenerate}) or $[i,i] = \{n\}$ (\textit{degenerate}) for some choice of $i \in \integers$. An \textit{elementary cube} is a finite product of elementary intervals $Q = I_1 \times \ldots \times I_n \subset \reals^n$, where $n$ is some fixed \textit{embedding dimension}. Its \textit{dimension} $\text{dim}\,Q$ is the number of non-degenerate intervals in the product. A \textit{cubical complex} $C$ is a subset of $\reals^n$ that is a union of elementary cubes. Specifying a field $\mathbf{F}$, we define $\mathbf{F}$-vector spaces $C_k = \{ \sum \alpha_i Q_i \mid Q_i \subseteq C \text{, } \text{dim}\,Q_i = k \text{, } \alpha_i \in \mathbf{F} \}$ for each $k \in \nats$, consisting of finite formal sums of elementary cubes. The \textit{boundary} of a non-degenerate elementary interval is given by the formal sum $\partial [i, i+1] = [i+1,i+1] - [i,i]$. For a degenerate elementary interval the boundary is zero. The boundary of an elementary cube $Q = (I_1 \times \ldots \times I_n)$ is a formal sum 
\begin{equation}
\label{eqn:boundary}
    \partial Q = \sum_{j=1}^n (-1)^{\sum_{i=1}^{j-1}\text{dim } Q_i} (I_1 \times \ldots \times \partial I_j \times \ldots \times I_n)
\end{equation} where we consider $\times$ as distributing over the formal summation. We can see that for $\text{dim}\,Q \geq 1$ we have $\text{dim}\,\partial Q = \text{dim}\,Q - 1$. Therefore we can extend $\partial$ to linear maps $\partial_k : C_k \rightarrow C_{k-1}$ via the mapping $\sum \alpha_i Q_i \mapsto \sum \alpha_i (\partial Q_i)$. Since $\partial \partial I = 0$ for any elementary interval $I$, we also see that $\partial_{k} \circ \partial_{k+1} = 0$ for all $k\in\nats$, so that $\text{im}\,\partial_{k+1} \subseteq \text{ker}\,\partial_{k}$. A sequence of linear maps
$$\ldots \rightarrow C_3 \xrightarrow{\partial_3} C_2 \xrightarrow{\partial_2} C_1 \xrightarrow{\partial_1} C_0 \xrightarrow{\partial_0} 0$$
with this property is called a \textit{chain complex}. The \textit{$k$\textsuperscript{th} cubical homology} of $C$ over $\mathbf{F}$ is defined to be the quotient vector space
$$H_k(C;\mathbf{F}) = \frac{\text{ker}\,\partial_k}{\text{im}\,\partial_{k+1}}.$$
This construction is functorial, i.e., given a suitable definition of a \textit{cubical map} $f: C \rightarrow D$ between cubical complexes, there is an induced map $f_k : H_k(C;\mathbf{F}) \rightarrow H_k(D;\mathbf{F})$ for each $k\in\nats$. We will not introduce these in general, but will note that given $C \subseteq D$, the inclusion map $C \xhookrightarrow{} D$ is cubical and hence induces maps on homology.

\begin{figure}
\centering
\includegraphics{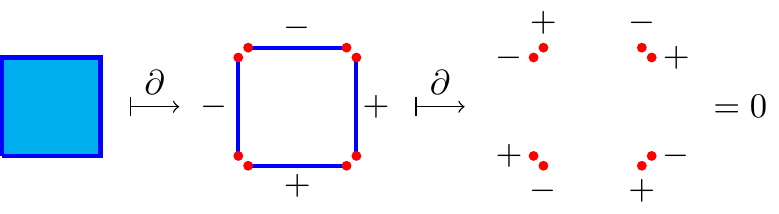}
\caption{Example of how the boundary operator $\partial$ acts on a simple cubical complex consisting of a single $2$-dimensional cube. Note how sum in equation \eqref{eqn:boundary} being alternating ensures that $\partial\partial = 0$.}
    \label{fig:cubical_boundary}
\end{figure}

\section{Histogram Reweighting}
\label{appendix:histogram_reweighting}
We make use of the correspondence between Euclidean field theory and statistical mechanics in order to apply histogram reweighting. This allows us to express the ensemble average of an observable $O$ at inverse coupling $\beta^\prime$ in terms of averages at any other $\beta$ according to the equation
\begin{equation}
\label{eqn:single_histogram_reweighting}
    \langle O \rangle_{\beta^\prime} = \frac{\langle O e^{-(\beta^\prime - \beta) S} \rangle_\beta}{\langle e^{-(\beta^\prime - \beta) S} \rangle_\beta}
\end{equation}
where $S$ is the action of the configuration (slightly redefined to pull the factor of $\beta$ outside) \cite{PhysRevLett.61.2635}. However, in practice we can only reweight so far, so that the sampled action distributions for $\beta$ and $\beta^\prime$ have a sizable overlap. To reliably extrapolate to a wider region we can make use of multiple histogram reweighing \cite{PhysRevLett.63.1195} where we sample at multiple inverse couplings $\beta_1, \ldots , \beta_R$. Suppose we sample $N_i$ configurations at $\beta_i$, then we can iterate the equation
$$e^{-f_{\beta}} = \sum_{i=1}^R \sum_{a=1}^{N_i} \frac{g_i^{-1}e^{-\beta S_i^a}}{\sum_{j=1}^R N_j g_j^{-1} e^{-\beta_j S_i^a + f_j}}$$
to estimate the "free energies" $f_{i} = f_{\beta_i}$ at $\beta_i$ up to an additive constant, where each $g_i$ is a quantity related to the integrated autocorrelation of the samples in run $i$. Given the $f_i$ we can estimate
$$\langle O \rangle_{\beta^\prime} = \sum_{i=1}^R \sum_{a=1}^{N_i} \frac{O_i^a g_i^{-1}e^{-\beta_k S_i^a + f_{\beta^\prime}}}{\sum_{j=1}^R N_j g_j^{-1} e^{-\beta_j S_i^a + f_j}}.$$

\section{Bootstrap Error Estimation}
\label{appendix:bootstrap}
In order to make any reasonable conclusions from the results of our analysis we need to be able to estimate the error in any numerical values obtained. While the error in ensemble averages can be directly estimated from the sample, we also calculate various fits to the data. The way in which error propagates here is not necessarily easy to calculate directly. Recall that the idea of bootstrap analysis is to sidestep these concerns by estimating the sampling distribution of a statistic directly. Suppose we obtain $N$ sampled configurations $S = \{ \boldsymbol\theta_1 , \ldots , \boldsymbol\theta_N \}$ and calculate some numerical statistic $f(S)$ from the data. Given some preset integer $N_B$, bootstrap analysis proceeds by:
\begin{enumerate}
    \item resampling $S$ with replacement $N_B$ times to obtain samples $S_1 ,\ldots, S_{N_B}$ each of size $N$; then
    \item computing $f(S_i)$ for each $i \in \{1,\ldots,N_B\}.$
\end{enumerate}
For large enough $N_B$, the distribution of the $f(S_i)$ approximates the sampling distribution of $f$ and we can estimate the standard error $$\sigma_f \approx \sqrt{\frac{1}{N_B - 1}\sum_i\big(f(S_i) - \overline{f(S_j)}\big)^2}.$$

\section{Distribution of $m_2$}
\label{appendix:m2_dist}

\begin{figure}[ht]
    \centering
    \scalebox{0.34}{\includegraphics{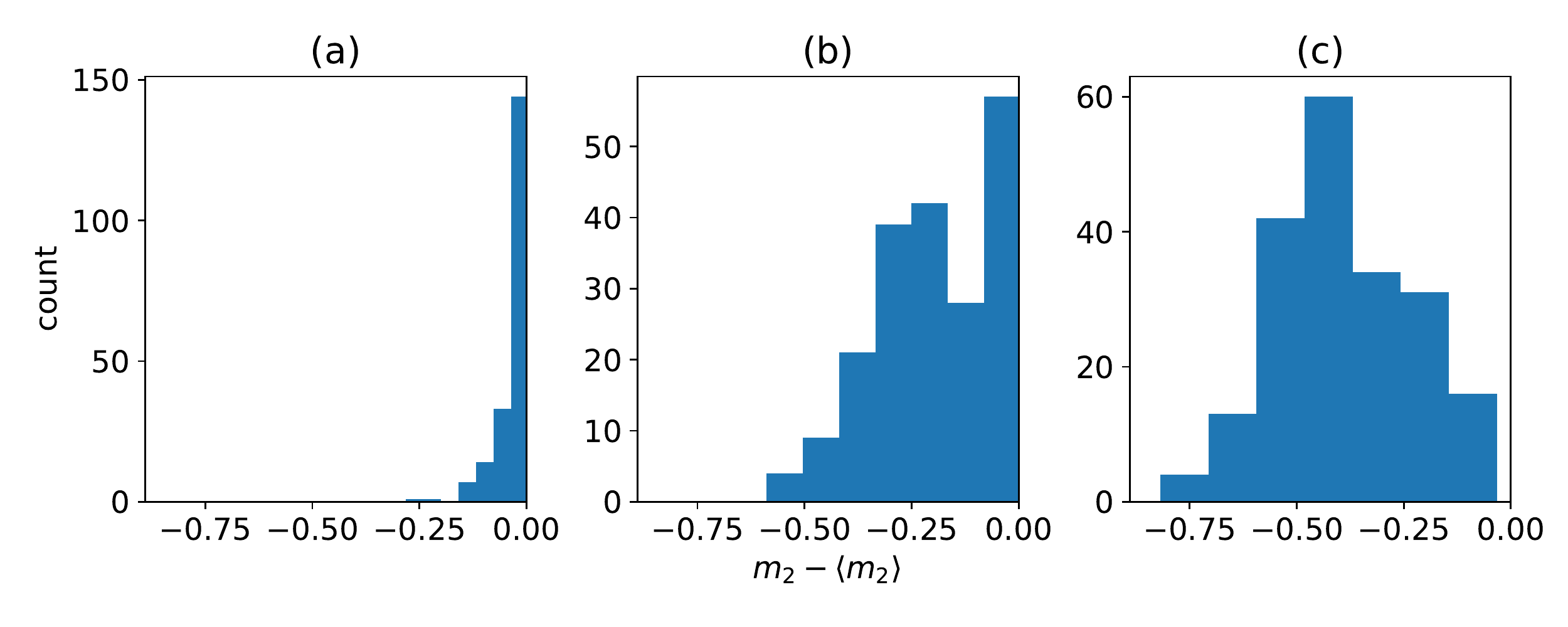}}
    \caption{The distribution of $m_2 - \langle m_2 \rangle$ with the twisted action at (a) $\beta = 2.2$, (b) $\beta = 2.5$, and (c) $\beta = 2.8$. Note that $\langle m_2 \rangle$ is the expectation as measured with the Wilson action.}
    \label{fig:m2_dists}
\end{figure}

Motivated to understand the error bars for the estimates of $\langle m_2 \rangle_{\text{twist}}$ in Figure \ref{fig:m2}, we look at the distribution of $m_2$ as measured using the twisted action on a $4 \times 20^3$ lattice in Figure \ref{fig:m2_dists}. We recenter the data using $\langle m_2 \rangle$ in order to compare the effect on $O_{m_2}$ at different values of $\beta$.

We see that at lower values of $\beta$ in the confined phase, the value of $m_2$ remains close to the average value measured with the untwisted Wilson action. As $\beta$ increases we observe a bimodal distribution, with some configurations maintaining an $m_2$ value close to the untwisted average and some joining a lower mode. A likely explanation for this behaviour is that for those configurations in the zero mode, the $H_2$ generator of the 4-torus responsible for $m_2$ does not correspond to the inserted vortex in the $yz$ plane, either because there is another vortex spanning a plane which is more easily observed or because the formation of a complete vortex surface along the twist in the filtered complex is being impeded, perhaps by intersection with other vortices. By the time we are firmly in the deconfined phase, the majority of configurations lie in the lower mode, signalling that the inserted vortex along the $yz$ plane is responsible for the value of $m_2$.

\begin{figure}[ht]
    \centering
    \scalebox{0.5}{\includegraphics{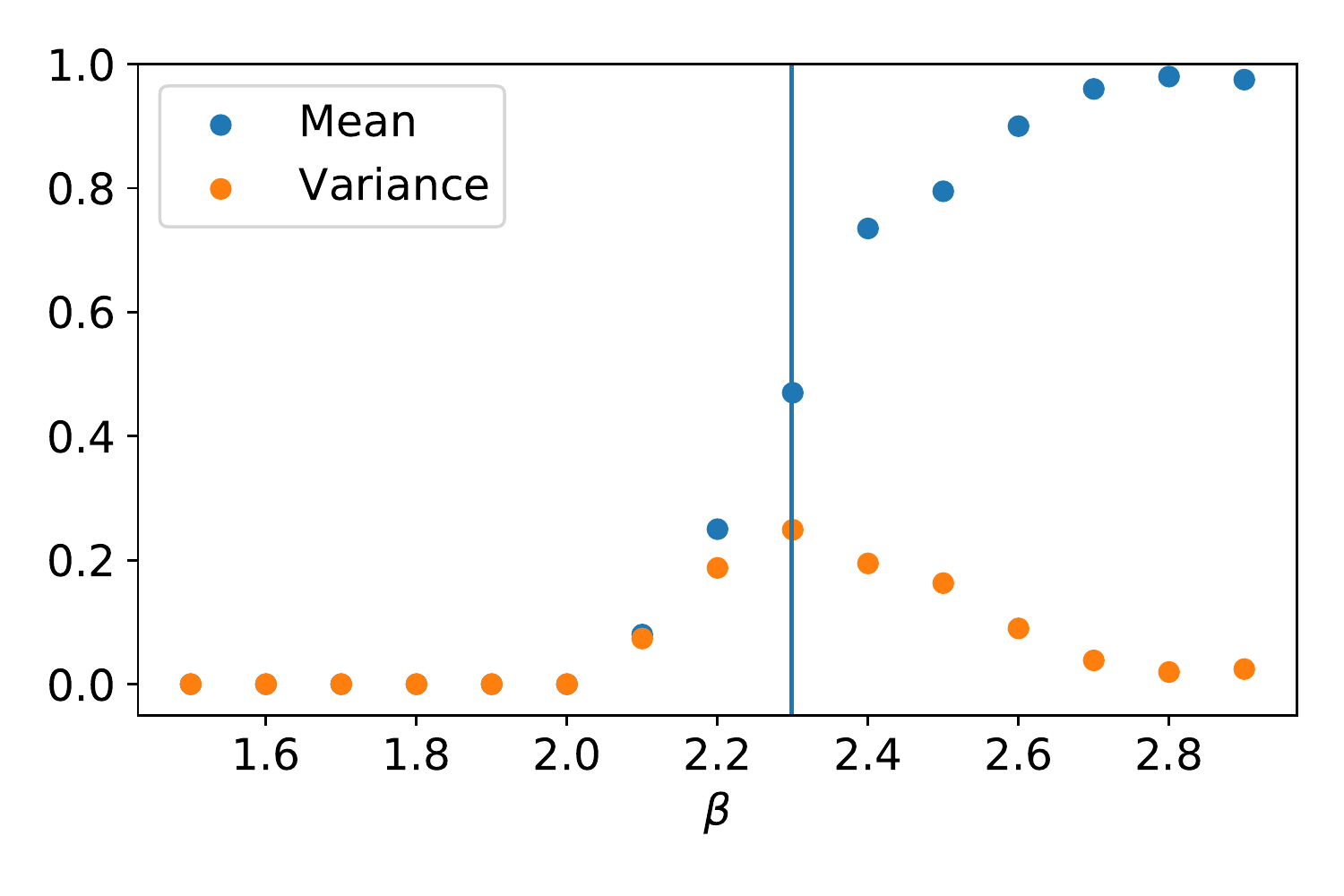}}
    \caption{The mean and variance of $I_{yz}$ as a function of $\beta$. The vertical line marks the location of $\beta_c$.}
    \label{fig:plane_proportion}
\end{figure}

To verify this picture, we can check if the $H_2$ generator responsible for $m_2$ is indeed represented by a $yz$ plane. We do this by recomputing the persistent homology but with the cubical complex only being periodic in the $y$ and $z$ directions and open in the $t$ and $x$ directions. This makes the final complex homeomorphic to a 2-torus which has a single $H_2$ generator: the $yz$ plane. Then we can simply check whether or not the birth time of this generator is the same as $m_2$. Denote by $I_{yz}$ the indicator function for a configuration that is $1$ if they match and $0$ if they do not. That is, $I_{yz}$ tells us if $m_2$ is determined by a $H_2$ generator which spans the $yz$ plane. The mean and variance of this indicator as a function of $\beta$ are shown in Figure \ref{fig:plane_proportion}.

We see that at low $\beta$, the inserted $yz$ vortex sheet is never responsible for the value of $m_2$. There are likely to be many vortices and those which span planes including the $t$ direction are smaller and likely to be picked up sooner in the persistence. Approaching and passing the transition point, the proportion of configurations for which $m_2$ describes the birth time of a $yz$ plane increases until close to $1$. There are fewer dynamically generated vortices and instead $m_2$ is determined by the one we inserted via the twisted boundary conditions. We note that peak in the variance of $I_{yz}$ gives us an alternative marker for the phase transition.


\bibliography{main}

\end{document}